# Survey on Wireless Information Energy Transfer (WIET) and Related Applications in 6G Internet of NanoThings (IoNT)

Pragati Sharma (SM-IEEE)[1], Rahul Jashvantbhai Pandya (SM-IEEE)[2], Sridhar Iyer (M-IEEE)[3], Anubhav Sharma[4]

[1]Dept. of ECE, S. D. College of Engineering & Technology, Muzaffarnagar, U.P., India-251001

[2]Dept. of EE, IIT Dharwad, Karnataka, India-580011

[3]Dept. of ECE, KLE Technological University-MSSCET, Belagavi Campus, Belagavi, Karnataka, India-590008

[4]Dept. of CS, LNMIIT, Jaipur,Rajasthan, India-302031

E-mail : sharmaprag79@gmail.com, rpandya@iitdh.ac.in, sridhariyer1983@klescet.ac.in, er.sharmaanubhav05@gmail.com

***Abstract*—** In the Wireless Information and Energy Transfer (WIET) technology, in addition to information, the electromagnetic waves carry energy in the energy harvesting mode, and hence, wiring infrastructure to charge the battery is not required. WIET is supposed to execute a vital role in the deployment and expansion of the 6G Internet of NanoThings (IoNT) devices which is envisioned to operate with limited-battery usage. As 6G technology will enable the use of wireless information for signaling information and energy transfer owing to the use of mm-wave/THz frequency for operation, antennas will be required at close proximity and hence, the Internet of Things/Internet of Everything/IoNT devices will be able to operate in near field region. In effect, the same electromagnetic wave will be able to carry significant energy to significantly charge the nano-devices.

This article contains an overview of WIET and the related applications in 6G IoNT. Specifically, to explore the following, we: (i) introduce the 6G network along with the implementation challenges, possible techniques, THz communication and related research challenges, (ii) focus on the WIET architecture, and different energy carrying code words for efficient charging through WIET, (iii) discuss IoNT with techniques proposed for communication of nano-devices, and (iv) conduct a detailed literature review to explore the implicational aspects of the WIET in the 6G nano-network. In addition, we also investigate the expected applications of WIET in the 6G IoNT based devices and discuss the WIET implementation challenges in 6G IoNT for the optimal use of the technology. Lastly, we overview the expected design challenges which may occur during the implementation process, and identify the key research challenges which require timely solutions and which are significant to spur further research in this challenging area. Overall, through this survey, we discuss the possibility to maximize the applications of WIET in 6G IoNT.

## I. INTRODUCTION

In the near future, the 6G technology is expected to integrate millions of devices with the internet due to seamless network capabilities. New dimensions will be explored for small, medium and large scale industries. In this entire portfolio, the new sector which is highly increasing is to minimize the dependency on wired devices for charging of any internet operated device. This is the need of wireless information and energy transfer (WIET) which uses the same electromagnetic wave for serving the communication purpose. Upcoming technologies in the field of wireless communication are expected to utilize the potential of Internet-of-Nanothings (IoNT) vision where many nano-meter wavelength devices are able to communicate and exchange information wirelessly. WIET allows to power up or charge the devices without the need of any wiring infrastructure.

Considering the mechanism used, WIET could be based on three classes viz., magnetic resonant coupling, inductive coupling or electromagnetic radiation [1]. The electromagnetic radiation, also known as radio frequency, enables WIET, and the radio frequency enabled WIET is able to cover long distances at low cost and is able to provide signal to the low power nano-devices. A low power battery operated device can be charged efficiently using the wireless energy without the need of a physical charging unit. The connectivity between the receiver and the transmitter can be divided between the following two modes [1]:

- **Mode-1** is the energy flow mode in which the wireless energy is harvested from transmitter to receiver. This energy harvesting (EH) is considered as a key component in upcoming generation IoNT devices since it provides 1) wireless remote charging which minimizes the maintenance and servicing needs of the internet-operated devices, and 2) significant enhancement in energy-efficiency.

- **Mode-2** is the information flow mode in which the information contents are transmitted or received through uplink or downlink communication, respectively. Multiple Input Multiple Output (MIMO) technology can be deployed to speed up the efficiency of mode 1 and mode 2 by the activation of multiple antennas at the transceiver which are used to enhance the data rate. It also maximizes the system efficiency, reduces the noise, and minimizes the power consumption so as to achieve the overall goal.

The size of the devices is becoming smaller day-by-day; however, it is able to fulfill more applications. The possibility of energizing the smart nano-devices seems to be possible after the reduction in power requirements of various smart electronic devices [2-4]. The concept imagines such high frequency radio waves which will carry both information and energy simultaneously from transmitter to receiver and will power up the low battery nano-devices. Sometimes, the term energy is replaced by power when a sufficient amount of energy is transferred at a high rate to meet the receiver battery requirements. The term wireless information transfer (WIT) is used for information/data flow between the transmitter and receiver in uplink and downlink mode targeting communication only, and wireless energy transfer (WET) for energy flow from the transmitter to receiver targeting energy transfer only. The design which is used for transfer of information and energy is WIET which attempts to accomplish a trade-off to communicate and energize the devices for the best use of radio frequency. Further, Nano-devices are small in size and need low power to remain functional and connected anytime and anywhere. The wireless network design needs to integrate newer changes to work on WIET which introduces new challenges and opportunities [4].

This survey is one of the first to discuss state of the art advances with a center of attention on aspects related to wireless energy transfer over nano-devices. Further, towards the end of the survey, several important current and future research challenges are presented followed by the proposal of the corresponding research directions. The main contributions of this study are as follows:

- Discussion and exploration of the state of art advancements towards obtaining solutions to the power related issues in the nano-devices through the use of WIET

- Devising a taxonomy, as shown in Fig. 1, of the WIET and related Applications in 6G IoNT survey.

- Presenting and discussing multiple research challenges of current and future research and the expected solution of these challenges with an outlook towards enhancing research in regard to finding the technology which better suits the 6G nanodevices communication and implementation on the low power IoNT devices.

The main focus of this article is to search out new dimensions of nano-devices and feasibility for the deployment of WIET in 6G IoNT. We have also attempted to cover all opportunities and challenges for sustainability of this research.

The rest of the article is structured as follows. Section II details the 6G network architecture and implementation challenges. In Section III, we describe the WIET architecture. Section IV focuses on IoNT, and the related design directions for the operation, In SectionV, we present a detailed literature review, and Section VI details the various applications of WIET technology in regard to the 6G nano-network and IoNT devices. In Section VII, we present the challenges related to wireless energy transfer in the 6G nano-network and devices, and Section VIII lists the future directions of research in regard to the existing challenges. Finally, Section IX concludes the article.

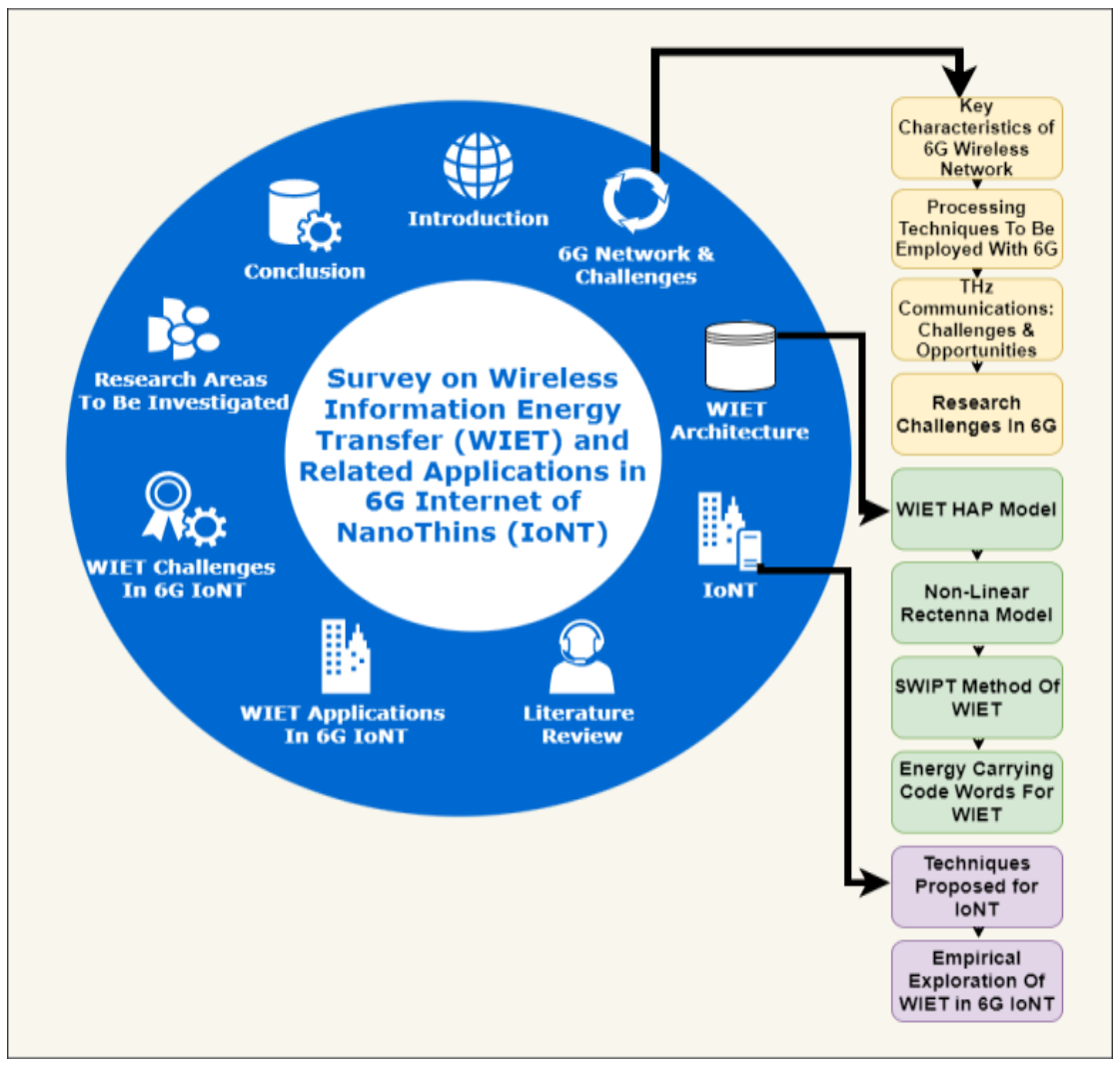

**Fig.1** Taxonomy of the WIET and related Applications in 6G IoNT survey.

## II. 6G Network and Implementation Challenges

6G offers extremely broad coverage and universal connectivity using artificial intelligence. It also opens doors to the possibilities for Space-Air-Ground-Underwater networks which consists of a joint layered tier network i.e., Satellite network, Air Network, Terrestrial Network and Underwater network. The communication occurs in the THz frequency band; thus, the wavelength (λ) is suitable for the nano-devices. The power requirement will also be low due to the very small size of nano-devices. The key parameters supporting 6G must provide services which are supposed to be implemented on ultra-high spectrum bands (0.1 to 10 THz) which travel shorter distances compared to the lower bands used in 4G and 5G. the potential technologies of 6G applications include IoNT, Molecular communications, Artificial Intelligence and Big data analytics, Laser communications and Visible light communications, Quantum communications and computing, Energy and spectrum efficient hardware and resource allocation holographic beam forming, Large intelligent surface, Dynamic network slicing, 3D networking, Unmanned aerial vehicle (UAV), Integration of access-Backhaul networks etc [5].

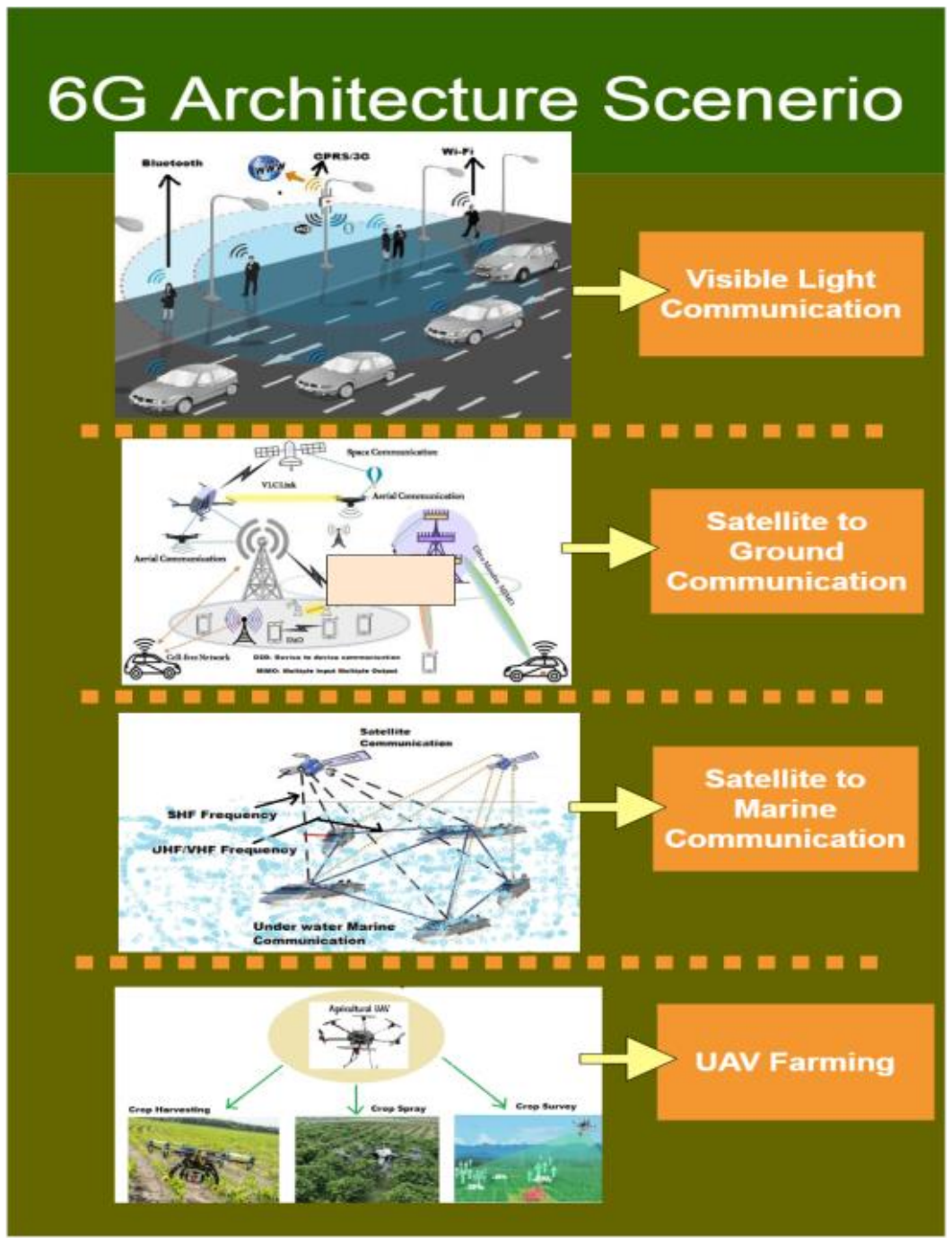

**Fig. 2** 6G Communication Architecture Framework.

The existing technologies are not comparable with the 6G technology. The bandwidth in sequence of 3G, 4G and 5G is 25MHz, 100 MHz and 1000MHz or 1GHz, Data rates are 14.7Mbps, 100 Mbps and 1Gbps, Latency is 100 ms, 50 ms and 5 ms respectively. However, 6G will unlock the potential of many applications providing more bandwidth, higher data rate and very small latency as shown in Fig. 3 [6].

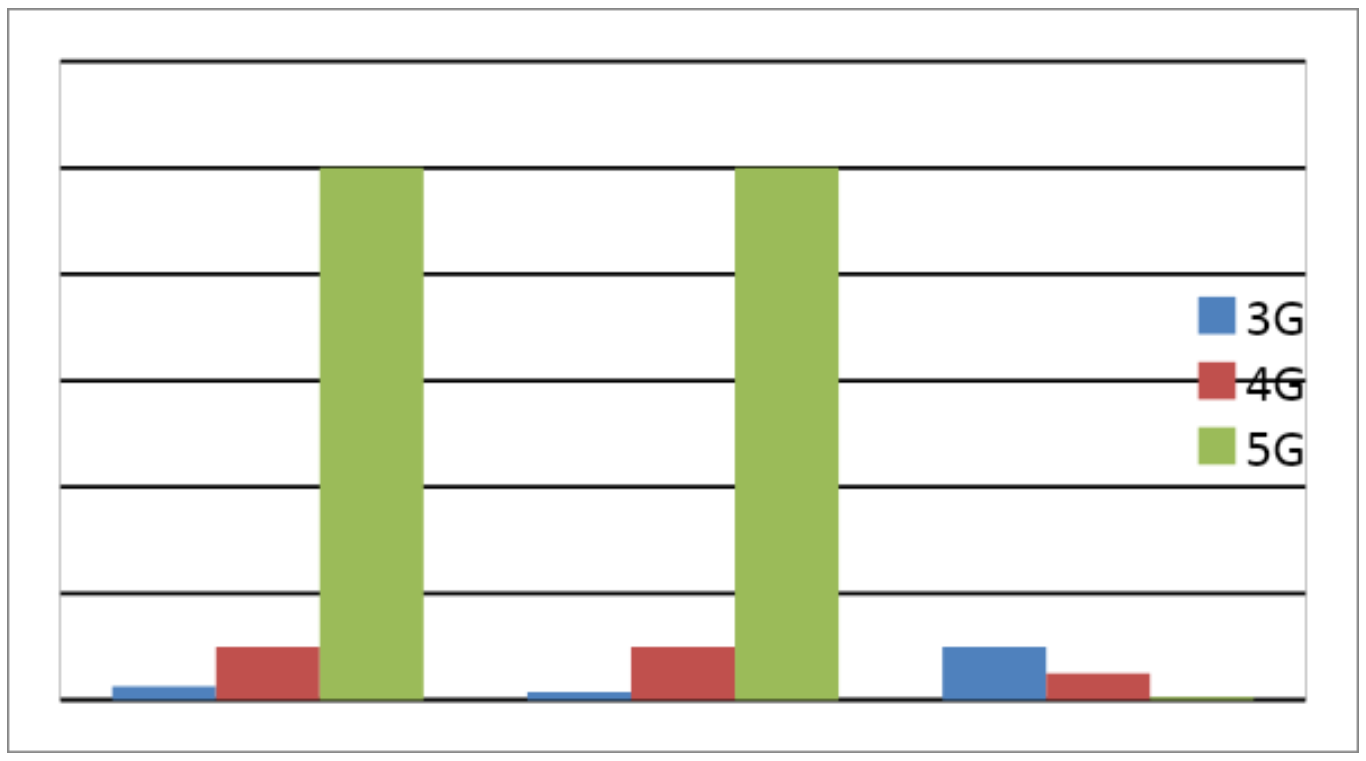

**Fig. 3** Comparison of bandwidth  data rate and latency for 3G to 5G technology.

## 1. Key characteristics of 6G wireless network [7]

- Peak data rates are of greater than 1Tb/s (nearby 100 times that offered by 5G)
- User data rates are of 1Gb/s (nearly 10 times that offered by 5G)
- THz frequency range: 0.1 – 10 THz
- Higher Spectrum efficiency that is nearly 5-10 times of 5G
- Boosted Energy efficiency that is nearly 10-100 times of 5G
- Ten times the connectivity density of 5G
- High throughput, network capacity
- Latency rate is  10-100 µs
- Mobility is high
- Enhanced data security
- Ubiquitous connectivity
- AI integrated communication

## 2. Processing Techniques to be employed in 6G [6, 8]

Processing techniques are those techniques which work on the concerned technology to make it best suited for the betterment of society in particular and public in general. 6G processing occurs at the frequency range of the THz band. 6G can employ the use of IoNT for connectivity and communication purposes when it deals with nano-devices. 6G is based on orbital angular momentum (OAM) multiplexing through which the transmission capacity can be increased dramatically. Another technique is Quantum communication and computing which is under investigation. Visible light communication and laser communication can be connected with 6G to widen the applications. 6G also supports artificial intelligence and blockchain based spectrum sharing to ensure the authenticity between end user and servers. It also allows the maximum use of remote resources. The throughput can be increased using non orthogonal multiple access (NOMA) which sends the signal at different power levels having the same carrier frequency. When NOMA is combined with mm wave frequency then it demonstrates more benefits to reduce interference and to improve the data rate. Another sub-domain of 6G is the super massive MIMO antenna which includes multiple antennas that provide  dramatic improvement in efficiency and offer super capacity.

## 3. THz communications: Challenges and Opportunities

THz band uses 0.1-10 THz frequency range which corresponds to 0.03-3 mm wavelength [7]. The immediate target for 6G is low THz band i.e, between 275 GHz to 3THz. The challenge of using higher frequencies for communication purpose are:

1 More towers will be needed to offer wide coverage.
2 The omnidirectional antennas are to be replaced with highly directional (pencil) antennas which are required to generate narrow beam widths.
3 The spectral efficiencies of the antennas need to be increased for effective communication. It can be achieved using super massive MIMO techniques.
4 Path loss becomes very high at higher frequencies therefore, the signal loses intensity over large distances and can be blocked by buildings.
5 Atmospheric absorption of high frequency waves occurs in addition to the natural Friis space loss of the electromagnetic waves at sea level with a variation of frequency under different humidity conditions.

At lower frequencies (below 6GHz), the attenuation is primarily caused due to molecular absorption in free space, which is of low amplitude; however, at higher frequencies (above 6GHz), the wavelength becomes micro/nano meter size. Such low wavelengths are the same as of rain, dust, snow etc., and hence, the effect of Mie scattering becomes more severe. The research shows that rain attenuation gets flatten out from 100 GHz to 500 GHz which shows that that rain will not cause any increase in attenuation at operating frequency above 100 GHz. The attenuation occurred due to rain at 1THz is 10 dB/km for moderate rainfall of 25 mm/h. This attenuation increases slightly, that is 4 dB/km more at 28 GHz frequency. The attenuation is controllable if the cell size is high, which is the case in all the mm-wave frequencies having urban cell sizes of nearly 200 m. Further, the rain attenuation can be minimized drastically at an increase in antenna gain which can be obtained by doing switching among antenna arrays and adaptive beam steering). Scattering, instead of diffraction, causes more notable changes at the nano-wavelength because of its lower size[9]. The opportunities which lie in the high frequency THz band is that it removes the spectrum dearth and capacity limitation which is the prime challenges in current wireless communication systems.

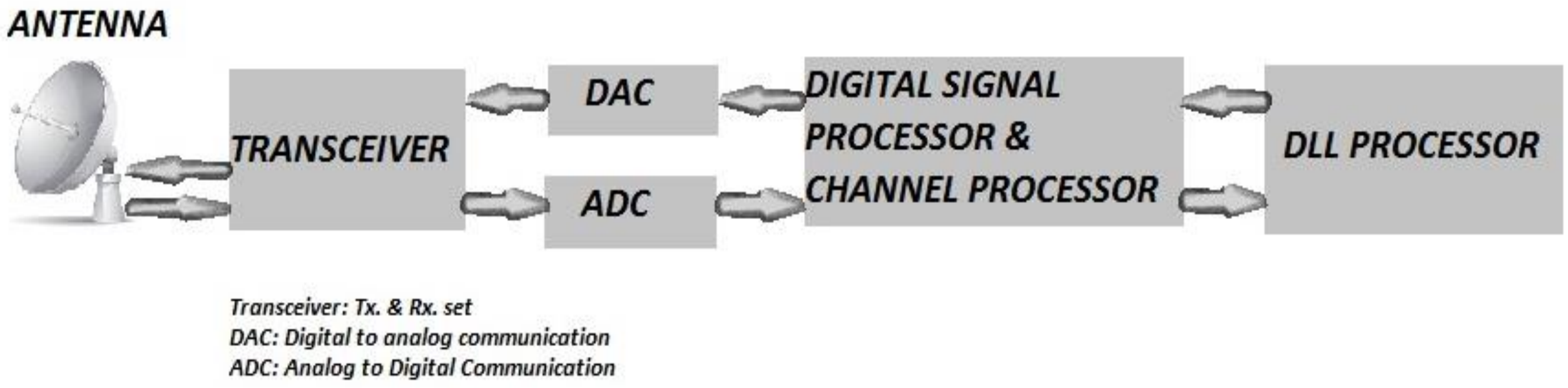

**Fig.4** THz wireless Radio Transceiver Hardware

Fig. 4 demonstrates the THz wireless radio transceiver hardware which is designed with a MIMO antenna array to receive the signals of low wavelength, and process these signals through the converters and processors to extract the information. Following aspects to be further investigated at THz communication operating frequency band:

- Transistor and all other active or passive devices with excellent high frequency characteristics
- Robust beam forming and scanning algorithm such as Hybrid Beamforming Approaches
- Low power consuming hardware
- Noise and Channel modeling
- Wireless charging
- Low complexity
- Energy efficient modulation techniques
- Ultra massive Multiple Input Multiple Output (MIMO) antenna system
- Powerful and high synchronization schemes
- low density channel codes

## 4. Research challenges in 6G

THz frequency band is expected to be used for 6G communication; however, it is not an easy task for wireless communication to operate over such a high frequency. Therefore, 6G faces many challenges before the successful execution over the THz frequency band. These challenges must be realized both in hardware and the communication aspects of 6G. Few of these challenges (demonstrated in Fig.5.) include [6]:

- Use of high frequency hardware components to reach the desired speed of operation
- Sufficient wireless energy transfer for speed charging of low wavelength and nano sized components
- Channel modeling and estimation for adept operation of high frequency
- Bandwidth enhancement for wide use and enormous capacity
- Use of VLC for minimization of Non-ionized radiation
- Intelligent dynamic spectrum access for virtual reality of spectrum availability to user
- Orthogonal multiple access for enhanced and efficient communication
- Focused directional networking for THz wireless communication
- High frequency operational directional base stations transceivers offer low distance coverage therefore large numbers of transceivers are required for 6G coverage.

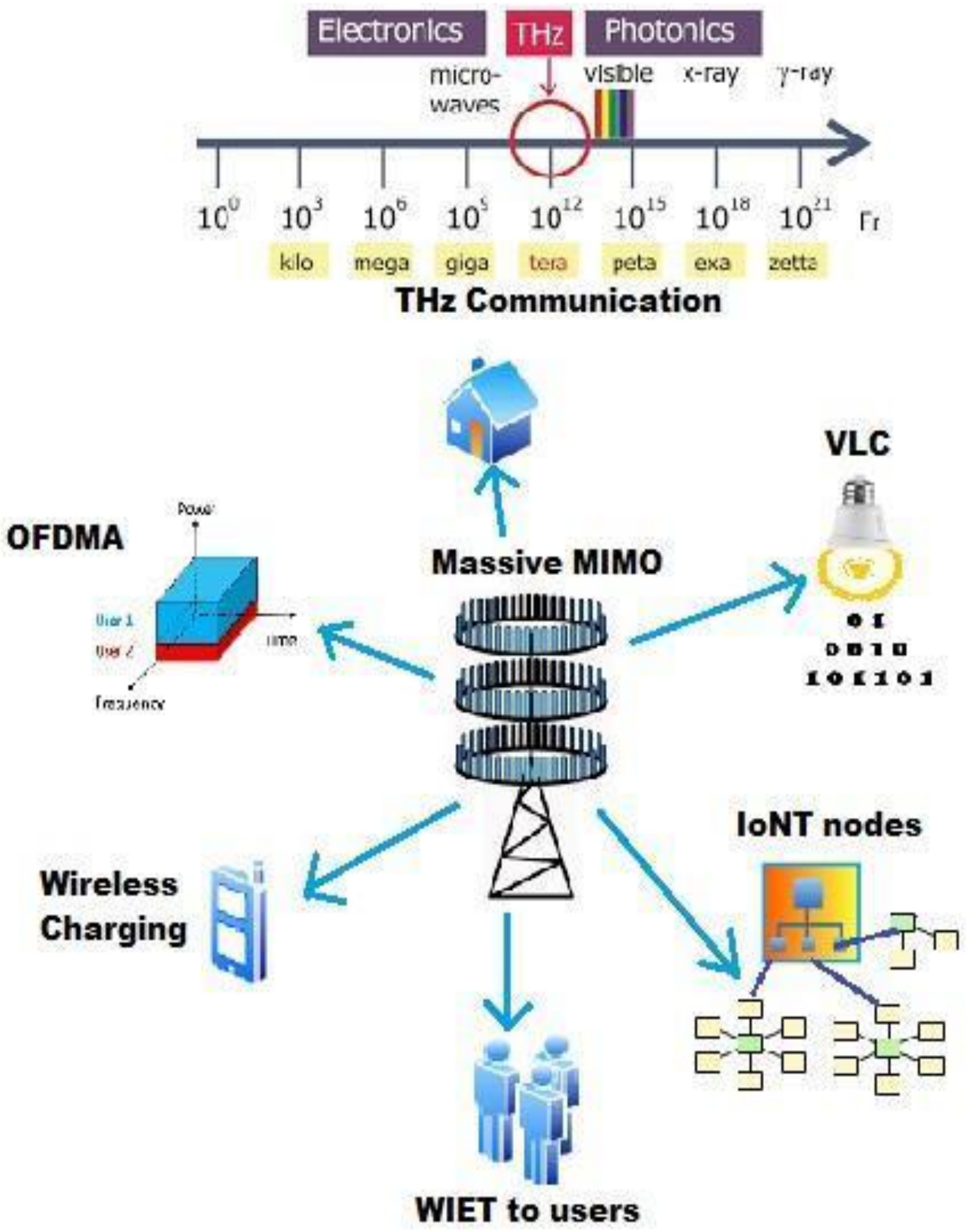

**Fig.5** The challenges in 6G Wireless technology.

Opportunities also arise in 6G communication along with the challenges. The operational frequency lies in the higher end of the frequency band, and hence, the wavelength is in the order of nanometer which allows the use of nano-sensors and other nano-devices with the 6G hardware. These nanosensors can be made operational on the Internet generating the IoNT network. The nano-devices face a problem of maintaining their power level because these are tiny devices and hence it becomes hard to power them up through a charger. The WIET technology is able to charge these nano devices. Information can be easily sent through the radio waves wirelessly; however, to send wireless energy over those waves, the receiver must not be far distant from the transmitter. Practical implementation of WIET in the 6G IoNT network is possible due to the lesser distance between the transmitting towers. Thus, all the receivers come under the Fraunhofer distance range which is also known as the radiating near field. The energy can be transferred to IoNT devices or 6G operational devices wirelessly without much effect of the attenuation. All IoNT devices are smart and able to communicate which makes them highly applicable in many fields like medical, health monitoring, commercial electronics, industry etc. WIET is the wireless transferred energy of low amplitude; however, it is sufficient to fulfill the power requirement of low power IoNT devices.

Apart from all the challenges, 6G also provides the opportunity to power up the receiver using radio frequency waves. For efficient charging of the device, the device should be a low power device with a small size. These parameters make 6G communication feasible and easier for nanodevices. The nanodevices are designed with the nanometer wavelength to operate on higher frequencies. Such devices are small in size and consume low power; therefore, a low energy harvest (EH) signal is able to charge the battery completely. When these nanodevices use the Internet for communication purposes, the new term is coined as Internet of nano Things (IoNT). All IoNT devices are smart and are able to communicate which makes them applicable in many fields such as, medical, health monitoring, commercial electronics, industry etc.

## III. WIET ARCHITECTURE

On the basis of radiation exposure, wireless power transfer may be of radiative type or non radiative type (Fig.6). The microwave and laser powers came under the radiative power zone whereas induction, magnetic, acoustic and capacitive powers came under non-radiative power zones [10]. The energy harvesting through wireless power can be implemented in direct mode or background mode. In direct mode the receiver directly receives energy from the transmitter which is actually meant to provide energy to the receiver while in background mode the energy is being retrieved by such device which is not aimed to power up the receiver directly but harvesting energy by some other process [11-13].

The radio frequency based WIET can be divided into three categories, as shown in Table 1. Further, it can be visualized through a model for transfer of wireless information and energy shown in Fig. 7.

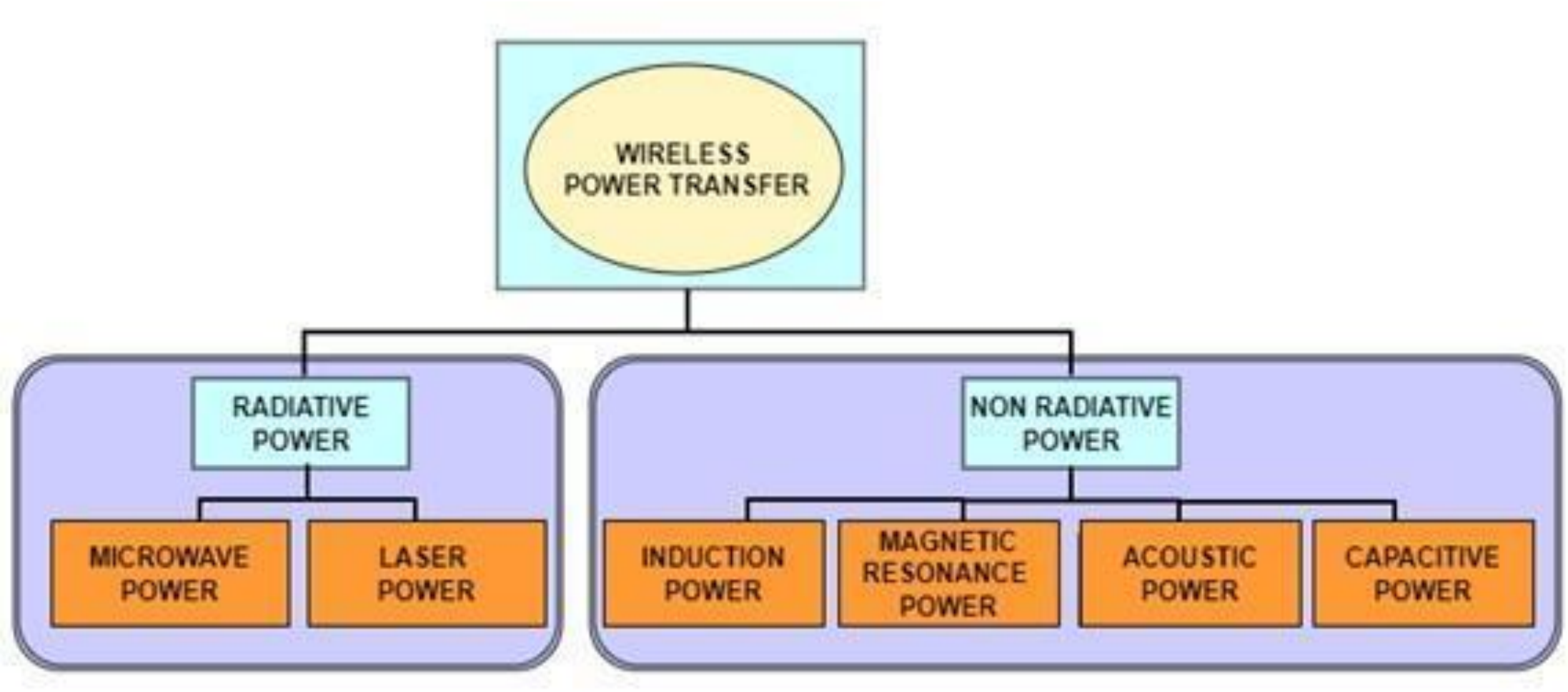

**Fig.6** Wireless power transfer categories.

**Table 1** Various modes of energy transfer.

| Parameter | Energy | Information |
|---|---|---|
| Wireless energy transfer (WET) | Transferred in downlink | -------------- |
| Simultaneous wireless information power transfer (SWIPT) | Transferred in downlink | Transferred in downlink |
| Wireless powered communication network (WPCN) | Transferred in downlink | Transferred in uplink |

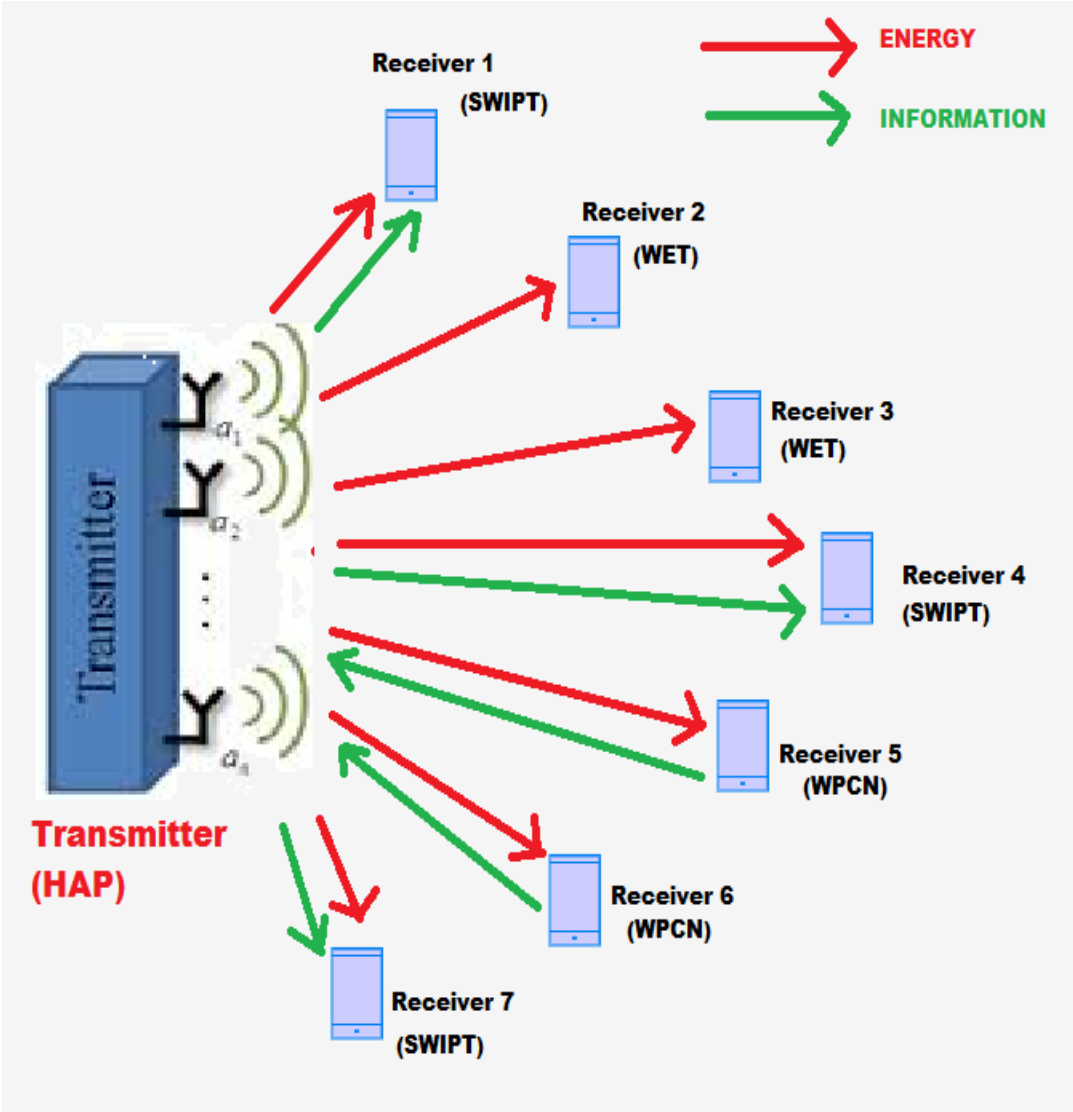

**Fig.7** Model for transfer of wireless information and energy

## 1. WIET HAP Model

The model in Fig.7 shows all the three modes of wireless information and energy transfer. Hybrid access point (HAP) is the transmitter used for transmission of information and possibly energy too. Energy flow is shown by the red line and information flow is shown by the green line. The energy transfer from HAP to Receivers is known as downlink, and the reverse process is known as uplink. Here, as shown in Fig. 7, Rx2 and Rx3 shows WET mode in which the energy is sent from HAP to the receiver. SWIPT mode is shown in receiver Rx1,Rx4 and Rx7. WPCN mode is shown in Rx5 and Rx6. Multiple antennas at HAP enable energy beamforming (EB), as shown in Fig.8(a), which is a technique used for efficient energy transfer. Multiple sharp beams in different directions are generated to minimize the near-far problem. It states that nearby users from HAP harvest more energy while the distant/far users harvest less energy[14, 15]. The high RF based wireless power transfer easily locates the receiver which is in the radiating near field zone as shown in Fig. 8(b). When the receiver is under the Fraunhofer's limit then the energy can be easily harvested because the receivers are easily diagnosed. Antenna array having a large number is desirable for the purpose of expanding the radiation in the near-field region. Radiating wireless power transfer allows an energy transmitter to charge multiple remote low power receivers or IoNT devices which are referred to as energy receivers. EH receiver is used to receive RF energy which is in the form of alternating current and needs to be converted in the form of direct current (DC) signals that are used to charge the battery of the low power receiver. In general, the rectifier circuit is converting AC into DC. Energy harvested rectified signal is received at diode output while energy harvested filtered signal is received at low pass filter output, shown in Fig.9.

## 2. Non-linear Rectenna model

The EH receiver receives the signal from the nano-antenna known as the receiving antenna or rectenna which is to be delivered to the diode for rectification and a low pass filter for smoothing the DC. The diode is a non-linear device which works in the forward bias only and converts ac signal into pulsating dc. This dc is then purified by a parallel capacitor forming a low pass filter (LPF) as shown in Fig. 9 [17].

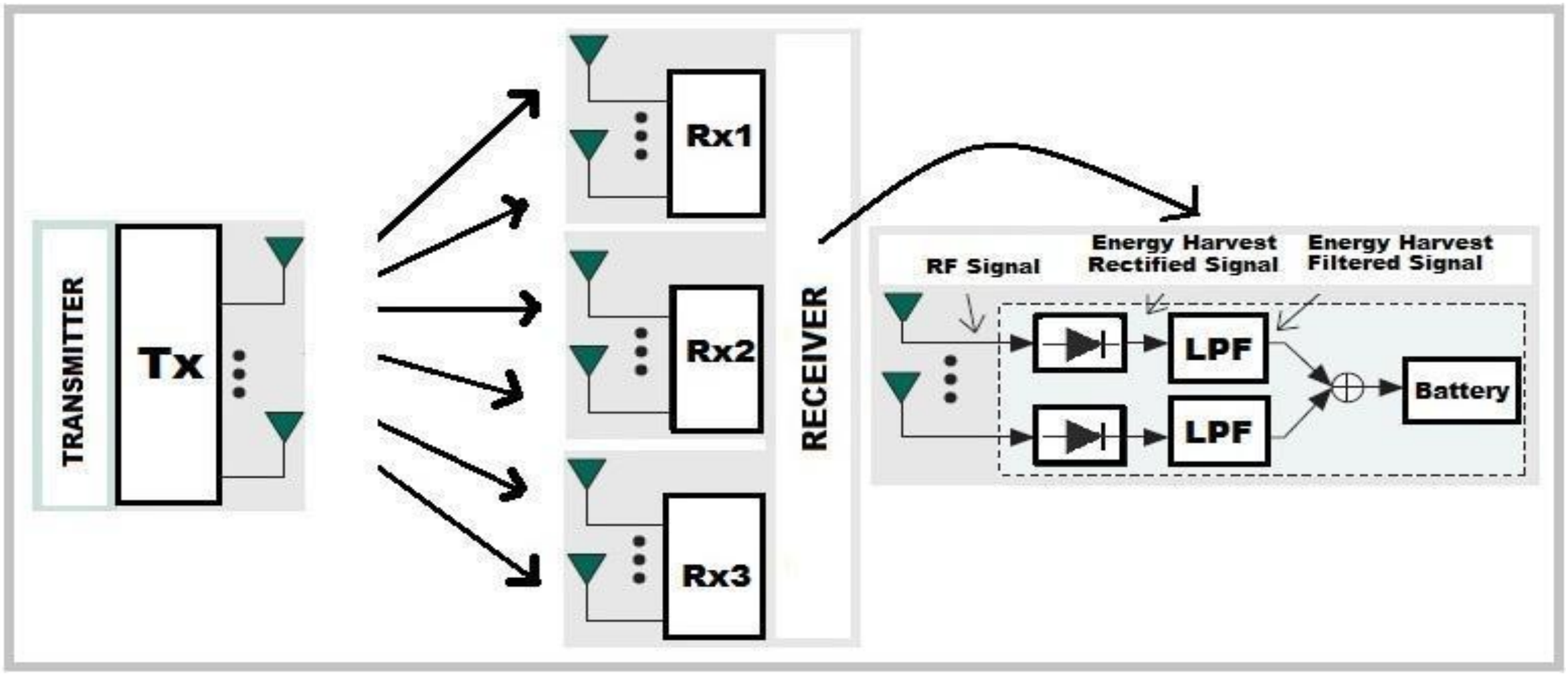

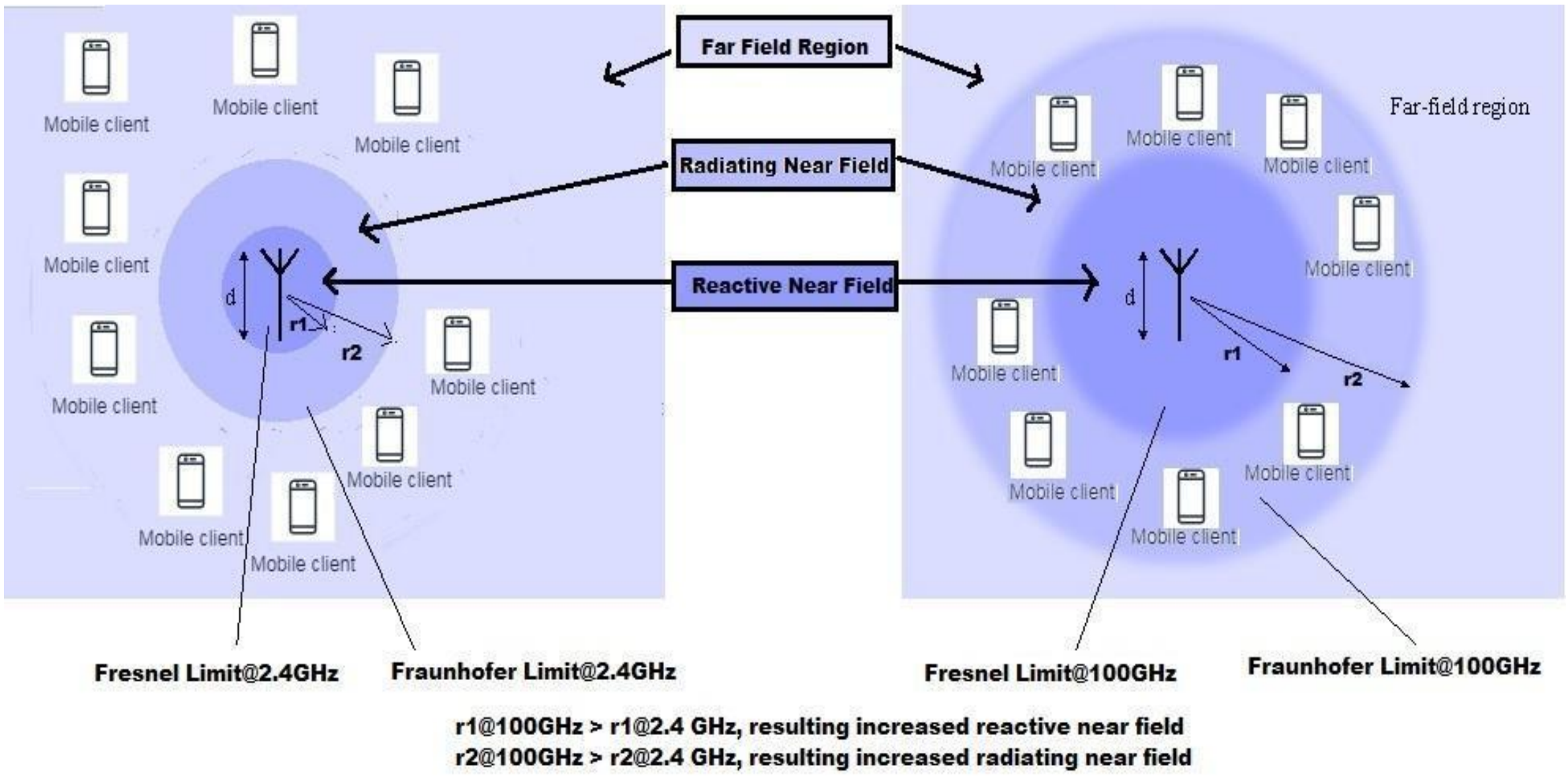

**Fig. 8** (a)WIET Infrastructure consisting of EH receiver [16] (b) Increased radiating near field at 100 GHz in comparison to 2.4 GHz.

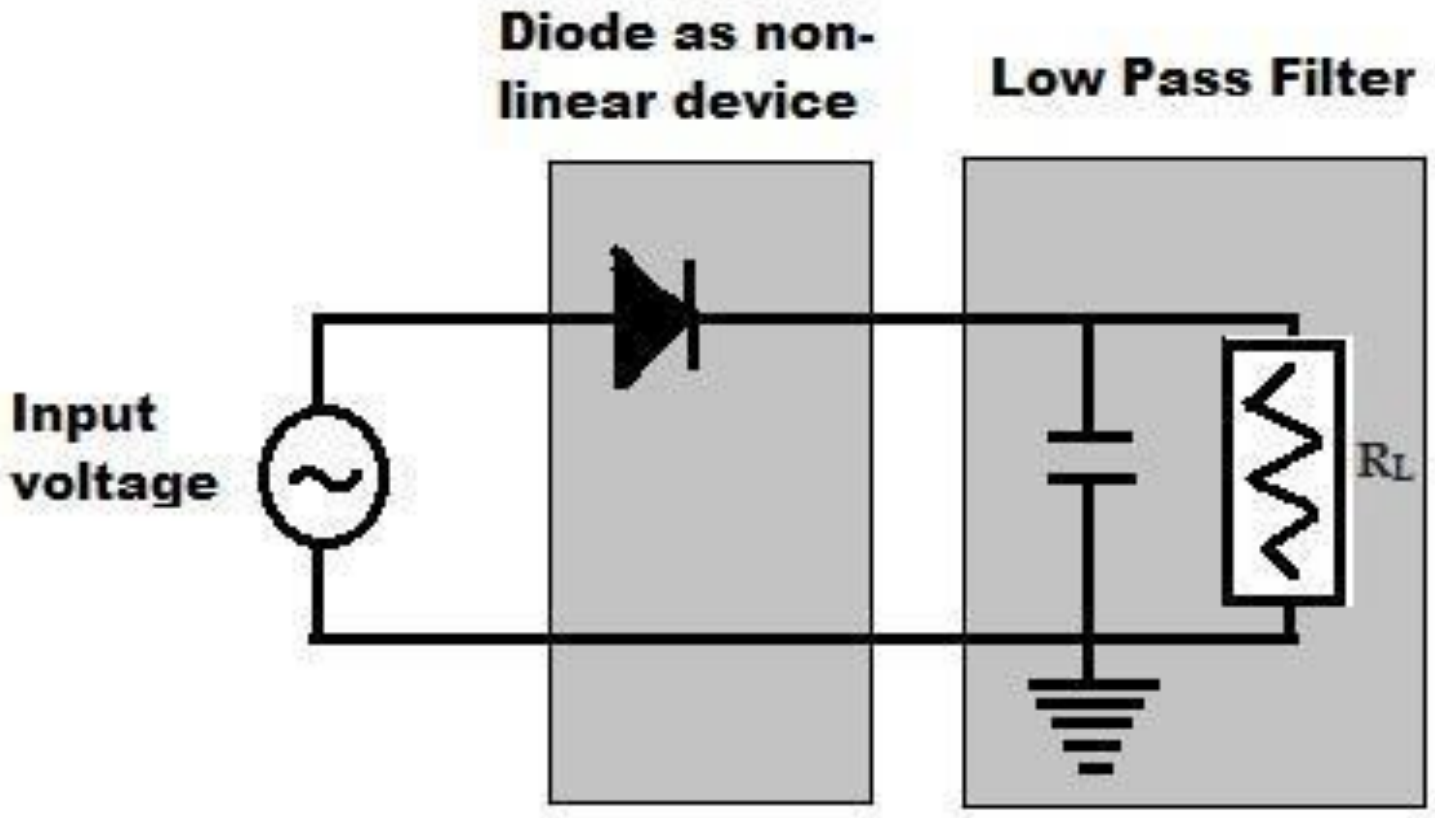

**Fig.9** Non-linear Rectenna model.

## 3. SWIPT method of WIET

When the paradigm of WIET is investigated to provide energy to the nano-receivers, it simultaneously transmits the energy with information over the spectrum. To avoid any interference which may occur due to energy transfer and may disrupt the information, a simple method is used which transmits the energy and information in orthogonal frequency channels. The SWIPT method provides a trade-off to balance the EH and information decoding (ID) shown in Fig.10 [18].

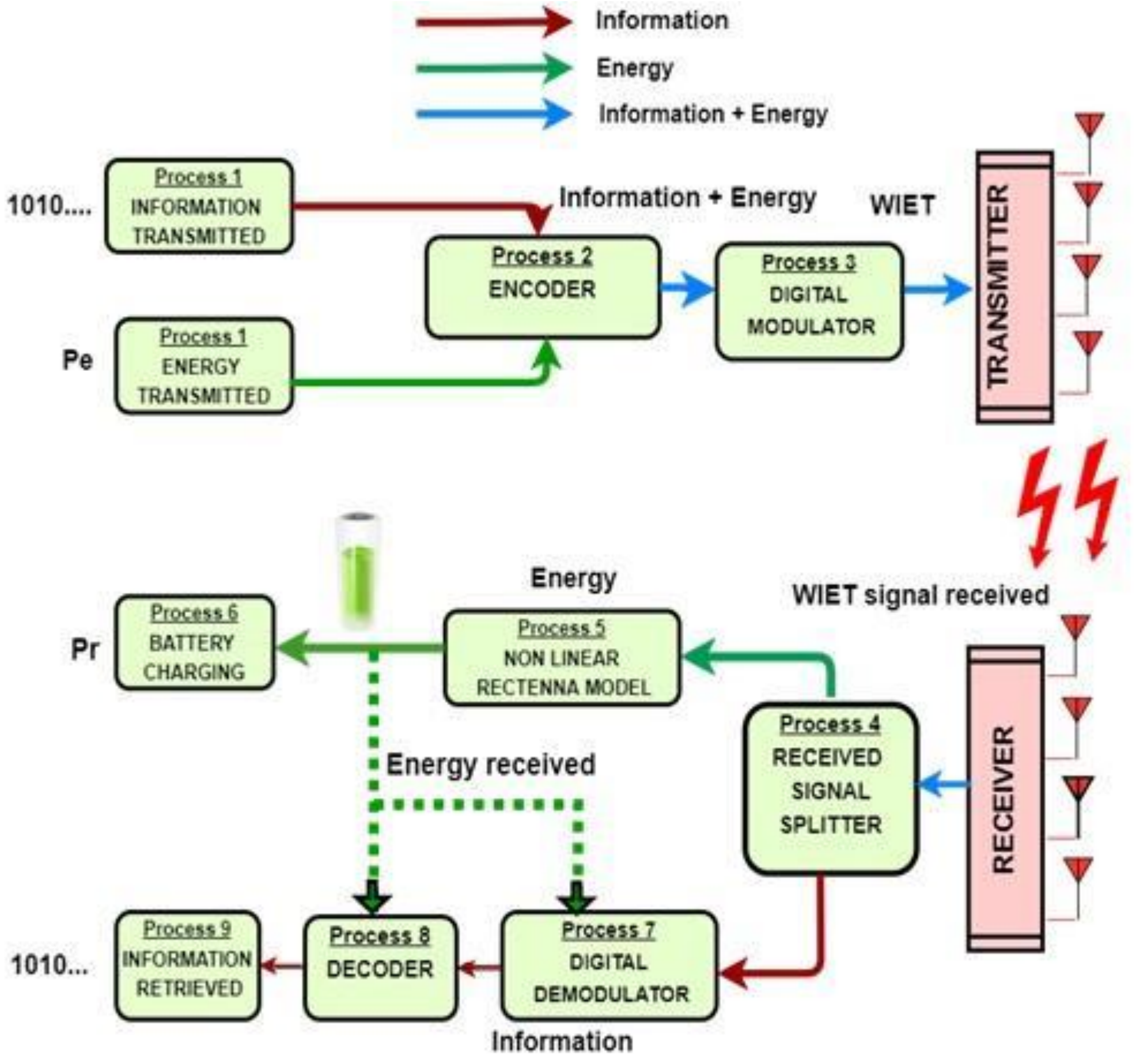

Fig10 - SWIPT method of wireless information and energy transfer [18]

## 4. Energy carrying code words for WIET

On the basis of different power requirements of battery capacity, the code words can be digitally modulated with a transmitted signal for wireless charging of the battery as shown in Fig.11. The low power receivers can be efficiently charged even at less harvested energy as they are coded with less number of 1's in data word. The energy is converted in terms of digital codewords using amplitude shift keying (On-Off keying) which is then transmitted using an energy transmitter. The codes are received and decoded at the receiver and used to power up the receiver [18]. The decoded codeword must contain the sufficient energy that should not be a cause of battery overflow or underflow condition.

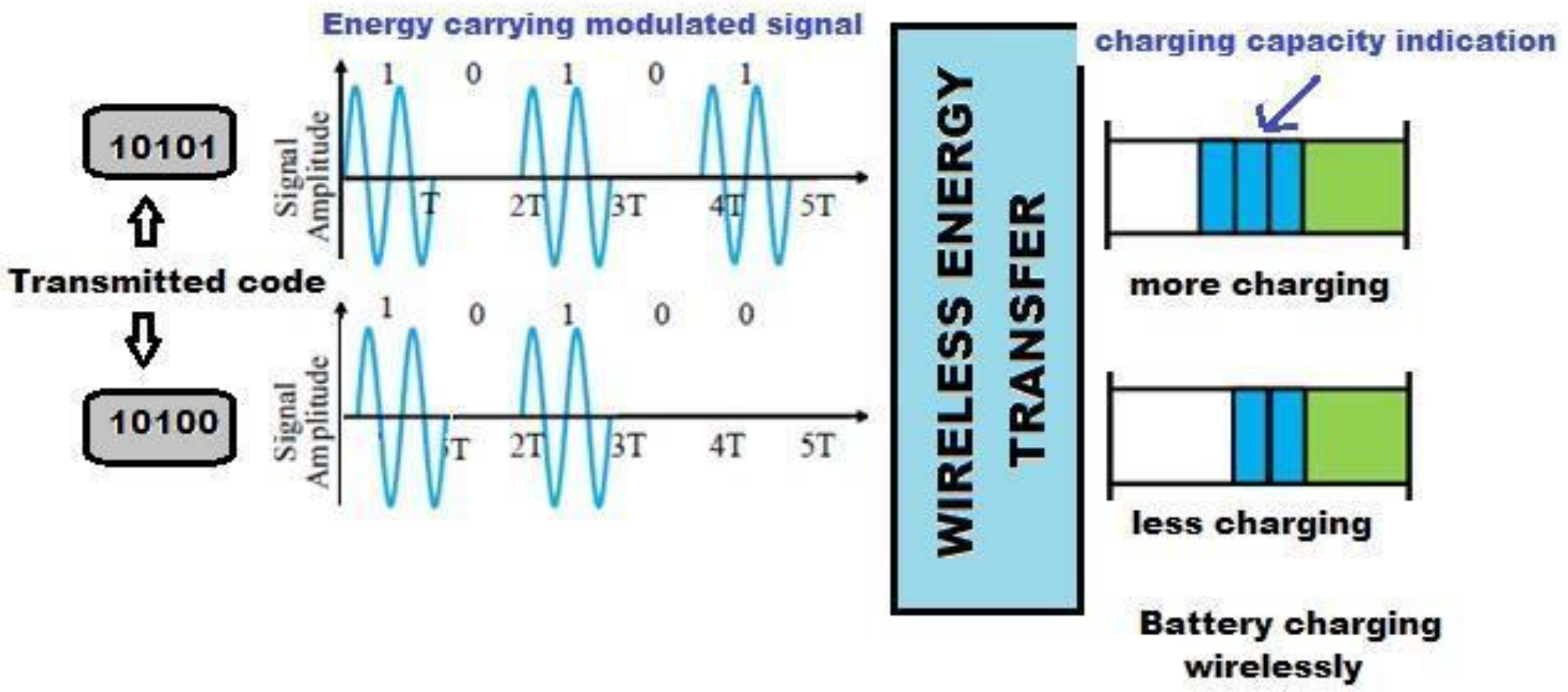

**Fig.11** Various energy carrying code words for WIET.

The energy received at the receiver becomes very low of the order of microwatt after passing through the atmosphere. Harvested energy of mm-wave is in the range of 1-5 $\mu$W when the source radiates 1-4 W power [19]. The authors in [20] observed that the received power level is not sufficient to charge wireless devices. The received power is in microwatt form so more focus should be given on the low-power devices such as small relays, nano sensors etc. Hardware development is required for widespread use of SWIPT technology.In the RF based WIET, various techniques can be adopted in the receivers to optimize the results. These techniques are time switching, antenna switching or power splitting. In antenna switching, receivers consist of separate antennas for information decoding or EH. The time switching mechanism is designed using one receiving antenna which is based on TDM for the operation. In power splitting, the power is divided in between information decoder or battery, as shown in Fig.12.

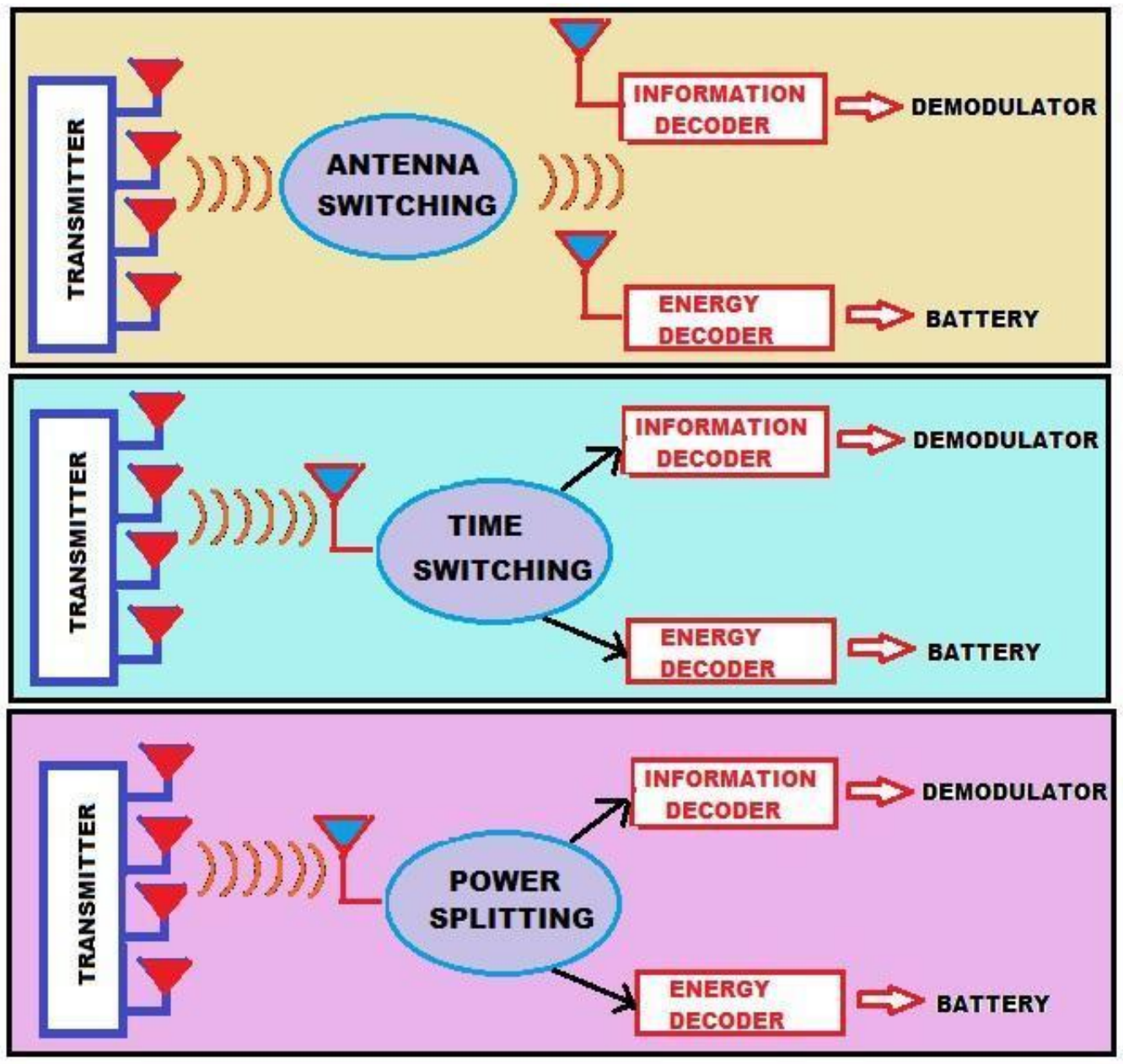

Fig.12 - Energy harvesting receiver designs [20].

## IV. IoNT

Internet of Nano things is a term which arises when nanomaterials, nano-implants and nano-biosensors are coined with internet of thing (IoT) networks. All the components to design the basic IoNT devices may include nano nodes, nano routers, nano micro interface devices and nano gateways. The IoNT basic building block which is interfaced at the nano-scale is shown in Fig.13 [21]. Nano nodes are used to sense the data and to process the information for further communication, and the nano routers perform the task of data collection from nodes and can control nodes for on/off. The collected information from the nano-routers is sent to the nano-micro interfaces. Gateways are connected with the nano-devices which transfer the information on the cloud through the Internet. Gateways are able to control the nanodevices remotely. The entire combination then becomes familiar to the IoNT.

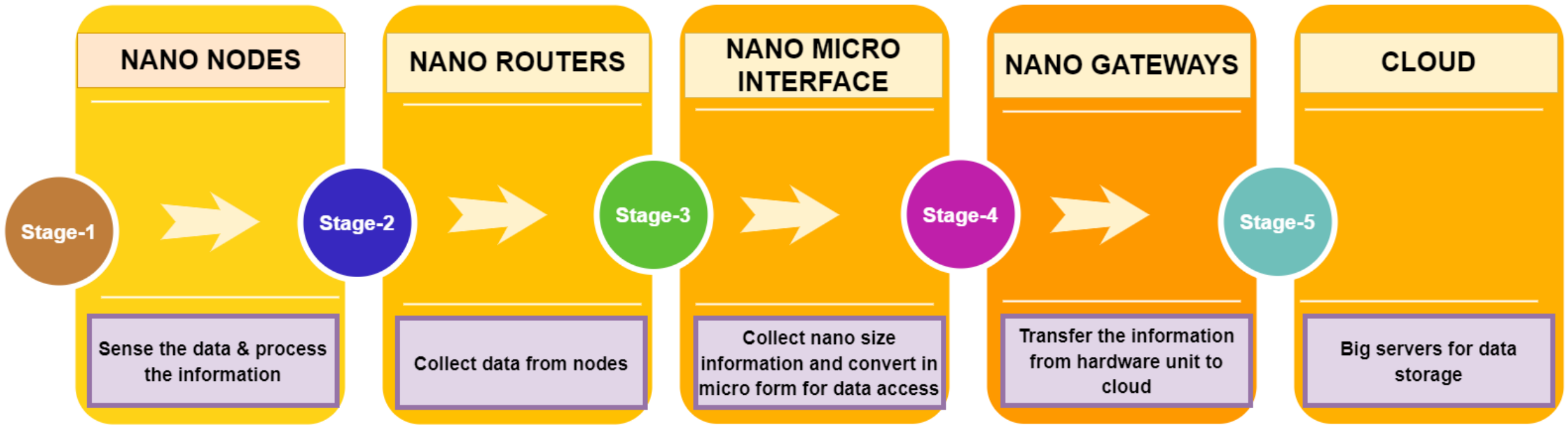

**Fig.13** IoNT basic building blocks

## 1. Techniques Proposed for IoNT

Two techniques proposed for communication of nanodevices are THz electromagnetic communication and molecular communication [22]. THz electromagnetic communication transmits and receives the information from nano transceivers through EM waves, whereas molecular communication transmits and receives the information which is encoded in molecules. Molecular communication is radiation-less and feasible even at such places where conventional wireless communication may create some health issues due to the use of EM waves [23]. This type of communication uses biochemical signals to transfer the information at nano-scale. IoNT is introduced from the concept of IoT with the only difference being in the size of things that may range from 1 to 100 nm. The 6G wireless communication needs to deploy nano-antenna to make the tiny devices communicable on the higher radio frequency. Graphene-based nano-antenna is suggested to be implanted in the IoNT technology because it has the ability to operate at the THz frequency band [24]. Nano-antenna and other basic nano-components are shown in Fig.14.

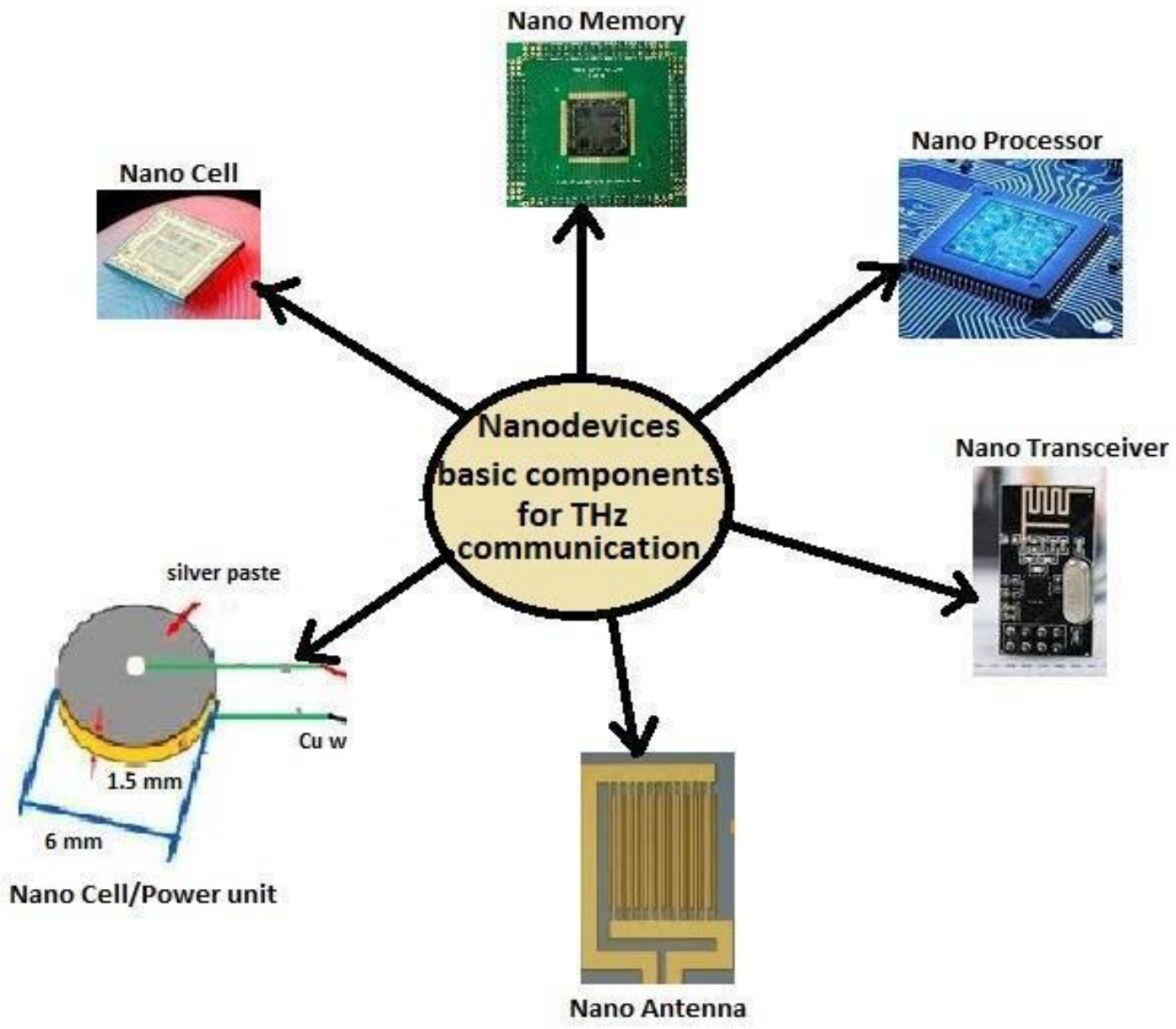

**Fig.14** Nanodevices basic components for THZ electromagnetic communication

The IoNT infrastructure can use other technologies such as cloud computing, WSN, big data etc. for collection, processing and refining of the data. The complete network of nano-devices is able to collect the data and granular information from places which are difficult to access with existing EMS/MEMS devices. Molecules are considered as the basic unit for bio nano-communication which works as an IoT sensor does. Internal parts of cells perform the task of various nanodevices like sensor, processor, actuator etc, and thus behave like a IoNT device, as shown in Fig. 15. Various cells are based on the propagation and reception of the molecules for the exchange of information between the cells [22].

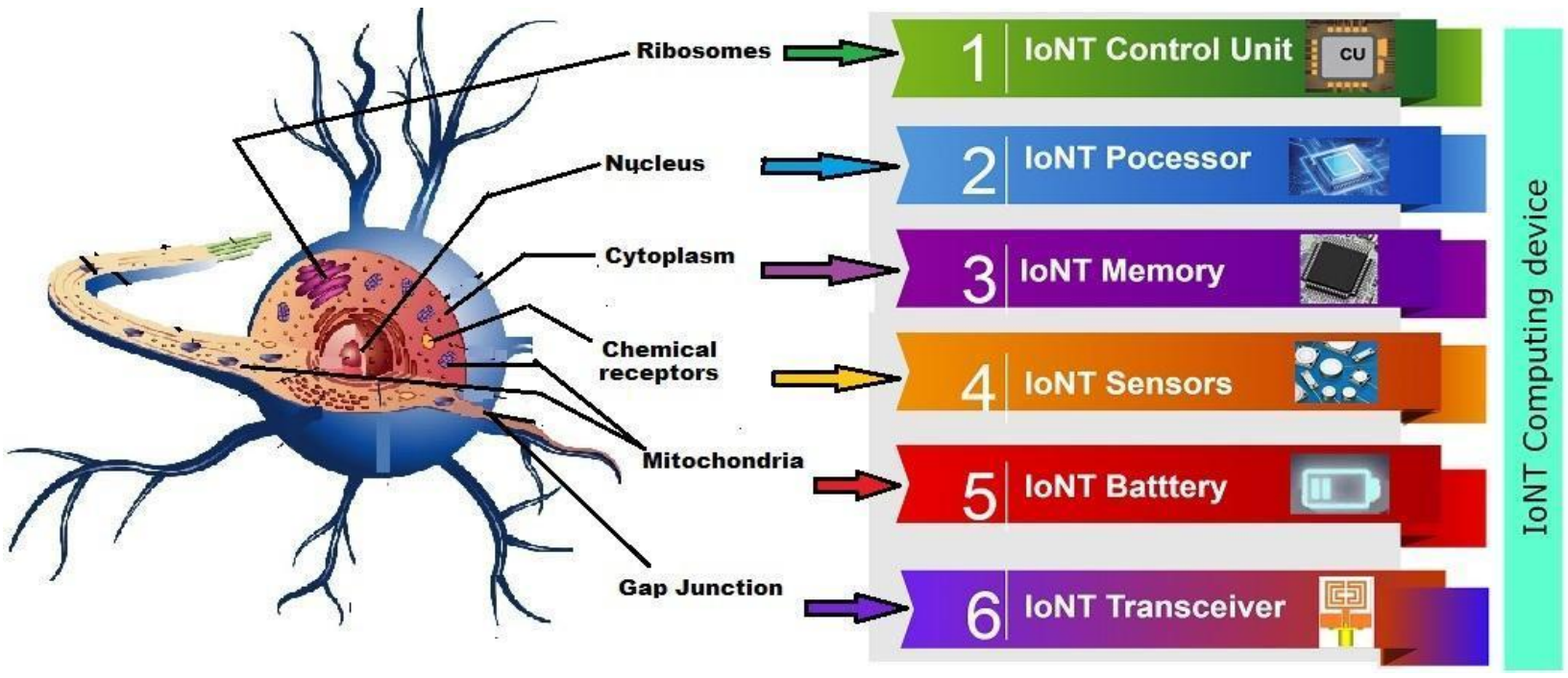

**Fig.15** Basics of Molecular communication.

## 2. Empirical Exploration of WIET in 6G IoNT

The nanodevices are able to work in the same fashion as a normal IoT device. These devices have their own transmitter and receiver for data communication, nanosensors for data collection, nano processors for data processing and sharing [25]. A small power unit is required to make all the building blocks operational. The main challenge is to provide an uninterrupted power supply which can be rectified using WIET. The role of 6G is interesting in the implementation of WIET in IoNT devices due to the fact that 6G operates at THz-frequency band having a wavelength of the order of nano-meters which can connect with a nano camera, nano sensor network, nano object or nano phone. The connectivity diagram of WIET in 6G IoNT is shown in Fig. 16 along with highlighted parameters or uses.

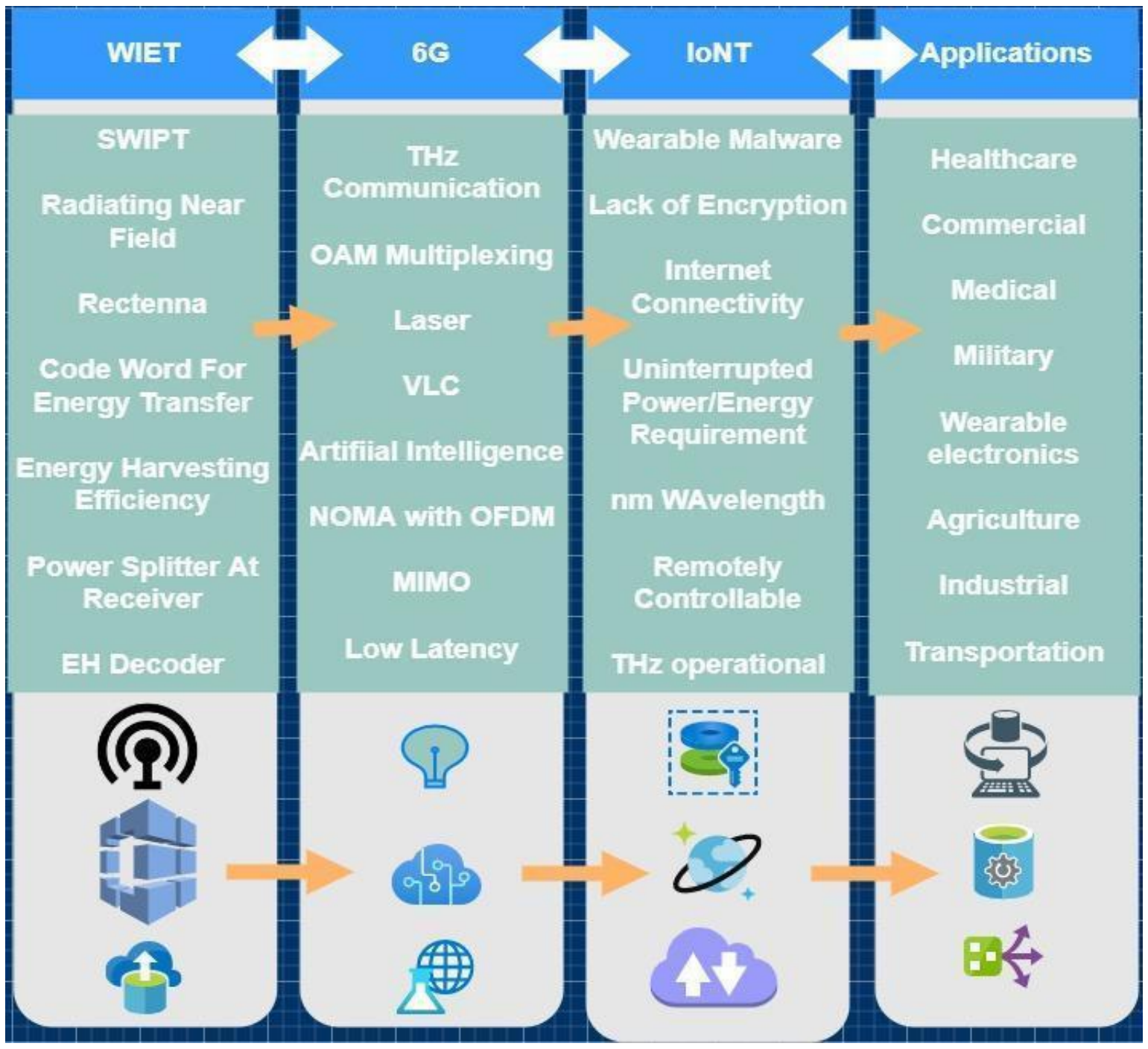

**Fig.16** An Empirical Exploration of the WIET in 6G IoNT

# V. LITERATURE REVIEW

In contrast to EH, several contributions and scopes are investigated in this section, which have been summarized in Table 1.

**Table 1** Literature Review.

| Reference | Aim | Problem addressed | Methodology | Key Results | Future Scope |
|---|---|---|---|---|---|
| [4] | Fundamentals of wireless information and power transfer | WIET | Paper explains various energy harvesting models through circuit simulation, experimentation and prototyping. The discussion on system design for WIPT is quite worthy. | Authors reviewed and discussed the network aim for 2 purposes of information and energy transfer. | Future work could be expended in new modulation formats for experimental validation, beam forming and modulation efficient networks. |
| [5] | AI-Based Joint Optimization of QoS and Security | 6G | Focus was given on the link-layer security in harvesting-based 6G IoT networks. The whole work is centered on EKF technique which is able to calculate the expected power needed to harvest with better accuracy. Authors have given the suggestion to evaluate the performance for network throughput in comparison with the commonly utilized service configuration method. | Authors suggested the use of adaptive security configuration methods which will be able to work on limited energy requirements. The suggested method is able to fulfill the need of energy requirements and capacity-constrained for IoT devices and has very low complexity. | Harvesting power can be predicted with high accuracy using EKF technique. This proposal can improve throughput rate and time. The directions were given to minimize the complexity in security configuration. |
| [6] | 6G Wireless Communications Networks | 6G | Paper depicts a detailed analysis from 1G to 5G. Paper elaborates a review on those technologies which are needed to increase energy efficiency and network connectivity. Paper also focuses on Blockchain technology and Quantum communication to increase security and privacy. | 6G is introduced to support such a massive internet of things.It will provide various advantages over the old technology like good intelligence and reliability, advanced security, adaptive data rates, ultra low latency, etc. | 6G will be able to provide seamless coverage without any communication difficulty and require new advanced modulation schemes, multiple access techniques, energy harvesting solutions, edge computing technique, integration of terrestrial and non-terrestrial communications, cell-free massive MIMO, blockchain and quantum technologies and adoption of artificial intelligence and machine learning techniques. |
| [7] | Optimal Resource Allocations | 6G | Authors worked on analyzing the THz-band overall interference. They also analyze channel capacity and channel dispersion functions through finite blocklength coding (FBC). | Paper focused on the development of resource allocation policies. The overall target is to maximize the capacity of the nano network. | Future scope found in Massive low latency communication that is preferred for 6G. It can prove beneficial to develop future IoT devices. |
| [8] | 6G Internet of Things | 6G | Authors made a comparison between 5G IoT and 6G IoT to elaborate network features and enabling technologies in Table III which guides about all the new features of 6G. | Author provides a taxonomy for 6G IoT applications and widens it for HIoT, VIoT, UAV, SIoT, IIoT etc. | Some promising challenges in 6G like security, privacy, hardware, energy efficiency etc. along with the possible direction of research have wide research scope in future. |
| [13] | Limitations of wireless power transfer technologies for mobile robots | WIET | Different types of mobile robots and their environments where they will be used are identified which is followed by power transfer scenarios. | New applications are searched out for mobile robots and WPT technology is reviewed. | Study of various types of wireless power transfer technologies makes it enabled to work for mid to far range communication. |
| [15] | Wireless powered communication | WIET | Methodology contains various aspects of design challenges, WPT technologies on real platforms and opportunities. | The design considerations of wireless powered communications were demonstrated through SWIPT and WPCN. | Research and innovations can be spur in the field of wireless power communication which will create a new dimension for wireless power enabling future devices to transmit information and energy simultaneously. |
| [16] | A Survey on Green 6G Network | 6G | Authors perform a near comparison between THz communication and VLC communication. | Authors found that 100 Gbps data rate could be achieved in the THZ communication while it is only 10 Gbps in VLC communication. The transmission power is also high in case of THz communication. | Likewise the VLC, in which light and high speed data can be transmitted the Green 6G communication, is emerging through which energy and high speed data can be transmitted sustainably. |
| [17] | Fading and Transmit Diversity in Wireless Power Transfer | WIET | The main focus is given to increase the efficiency of RF energy harvesters. Fading environment is analyzed for the purpose. | Authors developed A new design utilizing many dumb antennas to increase the RF energy harvesting at transmitter. | The role of multiple antennas may become beneficial in wireless power transfer even when there is no channel state information at transmitter. |
| [18] | Modulation and coding design for SWIPT | WIET | Paper focused on SWIPT and coding controlled SWIPT , MIMO aided modulation, SWIPT with modulation for single user and multi user. Paper shows a case study for 4-QAM,16-QAM and 64-QAM modulation design. | Paper establishes its work on coding part of simultaneous power transfer technology for one and more than one users. | Authors focused on three future challenges. These are concatenated code, coded modulation and adaptive modulation. |
| [20] | SWIPT in 5G mobile networks | WIET | Paper executes its survey for energy rate tradeoff and radio frequency allocation. | SWIPT technology is explored for the 5G mobile network. All the recent advances in this regard are summarized in paper. | Challenges still exist for high mobility users, ultra dense networks, artificial intelligence and information security are discussed. |
| [21] | Internet of nano things (IoNT) | IoNT | Paper suggests a clear picture of the components to design basic IoNT devices. | IoNT network may include various sensing nodes, processing units and gateways. IoNT basic building block interfaced at nanoscale . | IoNT will take IoT to a new level which worked on nanonodes and nano transceivers. Transmission and reception aspects of IoNT should be investigated deeply to maximize the efficiency. |
| [22] | IoNT future Prospects | IoNT | Basics of molecular communication is investigated. Paper suggests that the molecules work on chemical reactions for their actions therefore the electric energy requirement is quite low. Molecular communication can be adopted as a good method of communication. | Paper deeply concentrates on IoNT market trends, various applications , reasons to employ IoNT , challenges and Its nanosensor fabrication techniques. | Nanodevices' evolution and Internet of nano things have sufficient potential to connect every single object on the earth with the internet therefore future scope for IoNT is tremendous. |
| [23] | Green 6G Network | IoNT | Various promising technologies, THz band operation, VLC, molecular communication, quantum communication, blockchain for decentralized security are discussed. | Paper suggests molecular communication in a new communication paradigm. Advantages of molecular communication in comparison with radio communication are demonstrated.<br>Currently in 5G architecture, we are incapable of covering high altitude and deep sea scenarios. This drawback will be removed by 6G as 6G is able to provide wireless networks on a global scale through non-terrestrial networks. | Green 6G will explore many new aspects to increase efficiency and quality. In a true sense, it will be capable of providing anytime and anywhere a network. |
| [24] | Nanonetworks for communication | IoNT | Various deployment approaches are explained in a clear vision.The similarities between Micro robot nodes (nano network communication blocks) and nano network nodes (molecular communication blocks) are diagnosed. | Graphene is suggested as the future star of technology as it is capable of detecting extremely low concentration of charge. Grapheme can also be used in wearable sensors.<br>A clear overview of traditional communication and nano communication is explained. | Molecular communication scalability could be further investigated for short range and long range communication as it has a wide scope of research. |
| [25] | IoT, IoE and IoNT | IoNT | Difference of Internet of things, everything, nano things is executed making a base of people, data, process and things. | Authors predicted the challenges in IoNT that must be countered in channel modeling. | Paper tried to interface the nano devices with existing micro devices to utilize the existing infrastructure for future technology. More work needs to be done in this part. |
| [31] | Wireless powering IoT with UAVs | WIET | UAV-enabled Wireless Powering IoT technologies are identified and analyzed through simulation. | Simulation results found effective for Unmanned Aerial Vehicle enabling IoT technologies. | Ue-WPIoT have high potential in IoT . It can also be made functional for 6G to explore enormous applications in the micro and nano field region. |
| [32] | Mm Wave Textile Antenna for On-Body Energy Harvesting | WIET | Various parameters of the omni-directional mm wave textile antenna are analyzed for energy harvesting purposes measuring its impedance bandwidth and peak gain. | This work presents the first antenna on textile for wearable ambient RFEH in the 26 GHz and 28 GHz bands. The small antenna size is useful for wireless energy transfer easily. | The mm wave is the operating band of frequency so research can be explored in the direction to minimize induced losses at this frequency band. Textile based microstrip rectifiers can also be investigated for future research. |
| [33] | Intelligent Reflecting Surface WPC | WIET | Basic idea of an intelligent reflecting surface is taken to improve uplink and downlink energy and information transfer efficiency. IRS will work in between the HAP and user. | The results incorporate that as the reflecting elements are increased then the intelligent reflecting surface WPC rate will also increase. The phase shift is an important parameter for energy transfer and information transfer which influence the beamforming gain. | The IRS assisted WPCN has high scope to maximize energy from HAP to user and to transfer information from user to HAP. IRS can be proved as beneficial in increasing the energy efficiencies for wireless power transfer in 6G. |
| [34] | WPT in mmWave for rain attenuation | WIET | Paper suggests the wireless power transfer for coverage and exclusion radius in rain attenuation environment. | Paper confirms that harvested energy will increase at low coverage radius and vice versa. When WPT needs enhancement then small cells endowment can be used. | Rain attenuation is able to affect WPT highly so research is required for mm wave wireless power transmission. |
| [35] | H-IoT Emerging Technologies | IoNT | Keeping H-IoT centered, the role of AI, ML and Big date is discussed for various applications and circuitry of healthcare devices. | The Author delves into the ways nanosensors are transforming in the health sector. The term coined for this is H-IoT. | Some future research directions found prominent for research are Tactile Internet and Future IoNT. |
| [36] | Advancing Modern Healthcare With Nanotechnology, Nanobiosensors, and Internet of Nano Things | IoNT | Nano materials are classified on generations, size, dimension and chemical properties. Elements and general workflow of bio sensors are described to understand the H-IoT. | Paper explores the IoNT potential in the medical field. Authors also concentrate on health consequences and clinical risks that may occur during IoNT. | Study provides the future scope for targeted drug delivery, medical implants, breast implant, dental implant. |
| [49] | Massive Wireless Energy | 6G | Paper explained channel state information for simultaneous operation of a large number of devices. | Challenges and possible solutions for 6G WET technology are identified throughout this article and summarized. For sustainable WET, Conventional deployment to actively green WET is discussed. | Wireless energy transfer is a key enabler for upcoming generation. The use of nano devices may be beneficial to achieve the goal. |

## VI. WIET APPLICATIONS IN 6G IoNT

The use of sensors in multiple devices has increased enormously in the current decade after the proposed 6G scheme which gave rise to the invention of diverse micro/nanosensors. Varieties of applications are governed through the nanosensors at present and it is predicted to increase greatly in the near future. As, wavelength used in 6G communications lies in order of nanometers therefore the new components are in research which can operate on nano-meter wavelengths and are able to reduce the latency rate as well as respond to the wireless communication efficiently. The internet communicable nanosensors based devices are known as Internet of Nanothings (IoNT) which are pulling the attention of researchers due to their small size and low power requirements. WIET technology is found quite beneficial to sort the charging related issues of nanodevices and to communicate the information properly. IoNT devices are applicable in diverse fields like medical, commercial, environmental, industrial, research etc.[37-43].

1. **Health care Application**: These are used to analyze and examine human health, immune system support, radiation analysis, disease identification etc. The IoT system designed with the aim to focus on health aspects is popularly known as Healtcare-IoT or H-IoT. Normal fungal diseases to neurological diseases can be examined and identified through nano sensors. These nano sensors can work with IoT nodes properly when they will be powered through wireless energy without any disruption. Patient monitoring and alarm systems can also be incorporated with the use of artificial intelligence.

2. **Medical Application**: Applications of nanosensors are proposed for drug delivery and bio hybrid implants. The drugs are injected inside the body and are able to heal the body disease. This drug may be in the form of smart pills which consist of a complete set of transceivers in a tiny package. The power requirement of the smart pills is fulfilled by an energy device. Wired charging is practically not possible on such a small size transceivers but wireless charging through WIET can give it new dimensions and can remove the hurdles in medical applications. Bio hybrid implant is able to restore bone defects.

3. **Commercial Application**: WIET is useful in advanced mobile 6G communication. The wireless energy harvesting is beneficial in the smart city concept as it is able to charge nodes wirelessly. The applications are also found in security and surveillance.

4. **Military Application:** Wireless communication proved to be a boon in military applications. All the UAV vehicles are navigated through wireless communication and the utility gets increased when the wireless charging is also available. This can be done through nano-functionalized equipment.

5. **Environmental Application**: Nano sensor based 6G WIET architecture is expected to control air pollution and

biodegradation management in near future. The upcoming applications of WIET also include tracking of animals and biodiversity control.

6. **Industrial Applications**: 6G is expected to remain present “Everywhere” therefore the “Things” will be smart enough to maintain various quality controls of food, water, textile, health industry etc.

7. **Agriculture Application**: 6G WIET explores its scope in good quality seed identification., crop production and crop monitoring Soil moisture level can be monitored through. UAV can be used in seed disbursement.

The quality implementation of all these applications ensuring the long life of Nanodevices could be possible when these devices deploy WIET technology for communication and charging purposes. In the near future, numerous new sectors will also be able to take the quality benefit of 6G operated IoNT devices after solving the implementation challenges.

## VII. WIET IMPLEMENTATION CHALLENGES IN 6G IoNT

The 6G IoNT devices are small size, low power nano-devices which could be charged wirelessly through the 6G network. The prime issues which are required to pull the attention to work on 6G network are-

- **Network architecture**
  IoNT devices are ultra low power receiving architectures which need WET to fulfill their energy needs. In near future, when the transmission will start on THz band frequency in 6G then it becomes quite necessary to realize the potential of wireless energy transfer for nanodevices after IoT deployment. To maximize the harvested energy, the antenna should be highly directional allowing energy beamforming efficiently. Therefore there should be a deep focus on boosted efficiency, high gain HAP antenna design. The receiving antenna or rectenna should be like that which can operate on frequency selective bands to wide band spectrum.[44-48]. The rectenna hardware that can receive the maximum power efficiently at short wavelength waves is still a challenge for IoNT devices.
  The antenna can be designed using large array MIMO technique but the interaction of large arrays and high frequency gives rise to several implications during wireless transfer of energy because of the small size and low power handling capabilities of the receiver.

- **Mobility management**
  The efficiency of energy harvest systems decreases quickly when distances between transmitter and receiver are increased. This all happens due to the losses and attenuation occurred during the path. There is a big challenge to maintain the energy efficiency even when the communication is taking place at nanoscale.

- **Interference management**
  Another challenge exists with the transmission of such low wavelength signals which can counter several fading aspects during its pathway of carrying energy and information in it. Low wavelength signals may suffer from high scattering which can make the signal weaker and can be responsible for loss in energy and increased interference thus the efficiency at lower mm wave bands is tough to achieve.

- **Resource management**
  The near-far field problem may occur during the transmission of wireless energy to nanodevices in 6G because of the large Fraunhofer limit at GHz band. This should be significantly investigated to avoid any charging issue for wireless sensor networks, RFID tags or small nanodevices.

- **Security management**

The deployment of 6G WIET at nano-scale has to counter all the research challenges occurring in the field of WIET. Network security is a prime issue which cannot be compromised. Unauthorized access of data should be completely vanishing to maintain system security and to minimize risk[49].

## VIII. OPEN AREAS OF RESEARCH

In this section, the identified challenges are presented for open research to explore the future research directions to send energy wirelessly along with information. Various challenges are summarized in Table 2 and possible directions are suggested to obtain the solutions.

**Table 2.** Various existing challenges and possible directions.

| S.No. | Various Existing Challenges in research of WIET | Proposed Directions |
|---|---|---|
| 1. | WIET starts the device charging through energy harvesting from the transmitting station to mobile. What will happen if the receiving device discharges a little like 98%, whether the transmitter starts its charging? | It is important to establish a set of commands to charge the battery after a specific threshold limit. eg. if mobile discharges upto 30 %, only then the charging starts its functioning. |
| 2. | Whether 2G-4G operated devices are applicable to adopt wireless charging through some change in its hardware. An additional unit could be added to solve the issue. | Some additional hardware units could be designed to update the 3G-4G mobiles which can support the frequency transition. |
| 3. | What will happen if the technology upgrades further and the operating frequency gets changed? Whether it will have an impact on energy harvesting mode or WIET still remains unaffected. | The provision of further development should be accounted for in 6G phones so that it can support any change in frequency. |
| 4. | The quality of rectenna to be investigated to efficiently charge the receiver? | The size of the antenna depends on frequency. In general, it should be of quarter wavelength of the frequency used. The rectenna should be of narrow beamwidth and MIMO arrangement. |
| 5. | Whether the energy transmitter will charge idle devices too or can power other devices at finite time intervals or on-demand. | In the near future it is highly desired to monitor the battery status continuously by the transmitter. It can be implemented by sending one extra pilot pulse which will carry the battery status requirement. |
| 6. | What will be the transmit waveform design utilizing energy harvesting and focusing on energy beamforming for radiative near-field wireless power transfer systems? | Different nano antennas and their radiation pattern need to be investigated in labs or through simulation to understand waveform design and efficient pattern among all. |
| 7. | Research should be conducted to widen the design considerations of low power transceivers to encounter fading and interference effects. | All the indoor and outdoor models of wireless communication need to be investigated for nano antennas and high frequency operability along with rural/urban coverage. |
| 8. | The 6G technology must be investigated for radiation hazards (considering both, ionizing and non-ionizing radiation) before implementation. | The good side of 6G is that it can communicate on low power enabled transceivers. The low power impacts deeper radiation lesser inside the human body in spite of high operating frequency.<br><br>The Specific absorption rate (SAR) is the criteria to analyze the induced radiation inside the body which depends, SAR = $\sigma E_i^2/\rho$ where $\sigma$ is electrical conductivity , $\rho$ the tissue density of the body material and $E_i$ is the electric field inside that tissue.<br><br>The induced electric field increases with the increase in frequency while it decreases as the incident EMF decreases. |
| 9. | The attenuation occurring at THz frequencies is of due consideration for the success of WIET in the 6G network. | As suggested in Point 7, again all the indoor and outdoor models need to be examined against THz frequency and attenuation problems. |
| 10. | The IoNT technology may face many challenges during implementation.<br><br>● The first important challenge is that Nanodevices collect large volumes of confidential data, and concerns regarding privacy and security need to be addressed.<br><br>● Users of IoNT infrastructure need to be informed regarding who has access to their data and how their data will be used.<br><br>● Also, the collected data needs to be stored in a secure location with encryption and state-of-the-art cyber security protocols.<br><br>● If left unsecured, cyber criminals can illegally access this confidential data. In the case of a cyber-security attack, users may want to know who could be held responsible and which mitigation strategies can be executed.<br><br>Hence, IoNT developers need to consider these issues before the mass production and utilization of IoNT devices. | When wireless technology is connected with AI then numerous applications could be explored and a large number of problems could be countered.<br><br>IoNT infrastructure needs highly accurate transmission and receiving of low power data which when examined through all will result in less probability of security concerns. |

## IX. CONCLUSION

This article presents a detailed oversight of all pros and cons of the 6G communication and the implementation aspects on the low power IoNT devices. 6G will operate in the THz band which has numerous limitations. The possible 6G communication architecture is defined and various key possibilities, processing techniques, THz communication and research challenges in 6G implementation are discussed in detail. 6G communication better suits for transfer of energy along with the wireless information due to the working in the radiative near-field region which makes efficient transfer of energy possible. We envision the days in which tiny devices will be self-charged without getting connected with any hardware for charging, only needing 6G connection for charging. As the radio frequency lies in the THz band for 6G communication, more focus should be given for health issues also to minimize the radiation aspect in humans. Hence, in the article, we also suggest future research areas which need research activities for successful implementation of WIET in 6G IoNT as there may be some interference effects, fading, rectenna designing, orthogonal modulation etc.

Existing literature has been rigorously surveyed to further explore the WIET implementation aspects in low power devices like nanodevices needing very low power for operation which could be easily fulfilled through the transceiver giving due consideration to new modulation technologies like MIMO and OAM etc. Nanosensors based IoT devices have a wide application in the research field, commercial field & bio-medical field. The article also reveals the huge scope of WIET in Nanodevices to make these functional for a long time and provide long life so that all such devices which work on a limited battery need to energize wirelessly through the wireless power transfer. Till now, WIET is designed for the MHz band which is not powerful to charge the heavy battery devices; hence, the need is to increase the band frequency for decreasing the wavelength, increasing the radiative near field region, and to cut battery size as much as possible. This will make WIET suitable for charging low power devices. The ASK or other On-Off keying method could be utilized for transfer of energy in terms of code length. SWIPT model WIET has also been studied for finding the hardware needed during simultaneous transfer of energy with information.

Lastly, through this survey, the authors hope to inspire the use of wireless information energy transfer for successful implementation of 6G IoNT devices which will become a milestone for the future wireless communication.

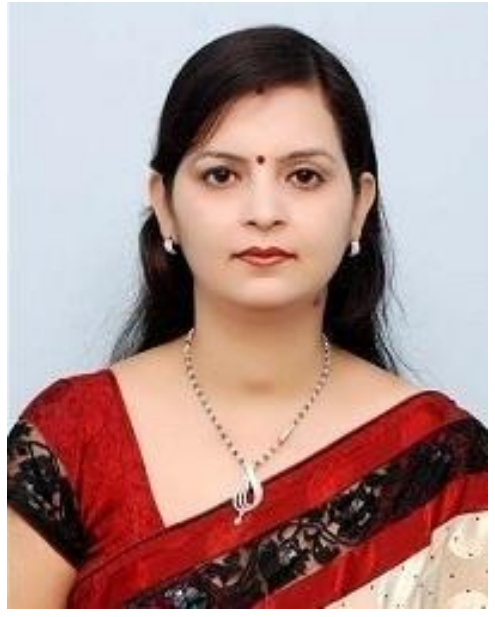

**Dr. Pragati Sharma** (Senior Member IEEE) received her bachelor’s degree in Electronics & Communication Engg. in 2002, Masters Degree in 2006 and Ph.D. degree in 2020. Her current research interests includes, electromagnetic radiation effects on human body, wireless communication latest technologies, 6G, Internet of Things, Nano technology and wireless information & energy transfer. She has published multiple technical books and reviewed several research articles. Currently, She is working as Head of Department, Electronics & Communication Engineering, S.D.College of Engineering & Technology, Muzaffarnagar,U.P., India.

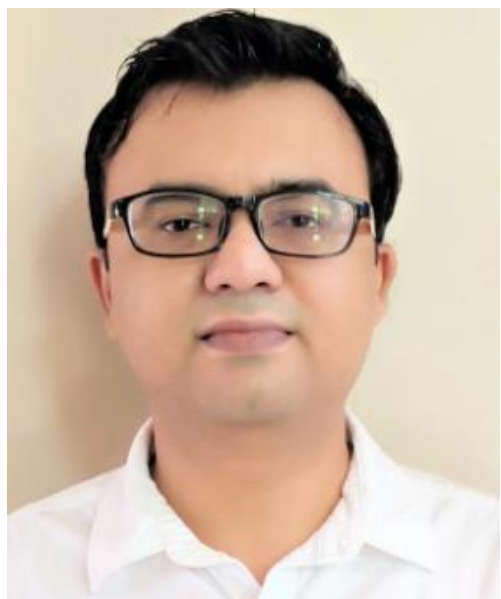

**Dr. Rahul Pandya** (Senior Member IEEE) completed his M.Tech. from the Electrical Engineering Department, Indian Institute of Technology, Delhi, New Delhi, in 2010. He completed his Ph.D. from Bharti School of Telecommunication, Indian Institute of Technology, Delhi, in 2014. He worked as the Sr. Network Design

Engineer in the Optical Networking Industry, Infinera Pvt. Ltd., Bangalore, from 2014 to 2018. Later, from 2018 to 2020, he worked as the Assistant Professor, ECE Department, National Institute of Technology, Warangal. Currently, he is working in the Electrical Engineering department at Indian Institute of Technology, Dharwad.
His research areas are Wireless Communication, Optical Communication, Optical Networks, Computer Networks, Machine Learning, and Artificial Intelligence.

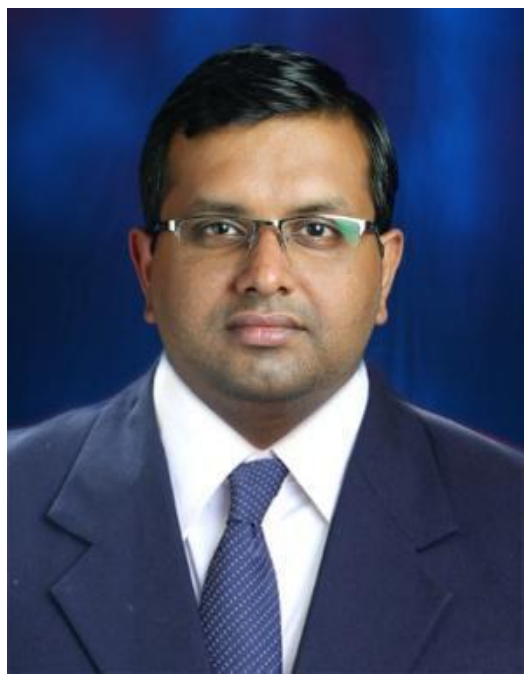

**Dr. Sridhar Iyer** (Member IEEE) received the M.S. degree in Electrical Engineering from New Mexico State University, U.S.A in 2008, and the Ph.D. degree from Delhi University, India in 2017. He received the young scientist award from the Department of Science and Technology, SERB, Govt. of India in 2013, and Young Researcher Award from Institute of Scholars in 2021. He is the Recipient of the 'Protsahan Award' from IEEE ComSoc, Bangalore as a recognition to his contributions towards paper published/tutorial offered in recognized conferences / journals (during Jan 2020 - Sep 2021). He has completed two funded research projects as the Principal Investigator, and is currently involved in on-going funded research projects as the Principal Investigator.

Dr. Iyer serves on the review panel of high-impact Journals such as IoT Magazine, IEEE, CCN, Elsevier, PNET, Springer, etc, and is on the Editorial Board of the Journal of Computer Science, Sci-Hall, Canada. His current research focus includes semantic communications and spectrum enhancement techniques for 6G wireless networks, and efficient design and resource optimization of the flexi-grid EONs enabled by SDM. He has published over 90 reviewed articles and multiple book chapters in the aforementioned areas.

Currently, Dr. Iyer serves as an Associate Professor in the Dept. of ECE, KLE Technological University-MSSCET, Belagavi Campus, Belagavi, Karnataka, India.
**Google Scholar**: https://scholar.google.co.in/citations?user=2hbORHEAAAAJ&hl=en;
**ORCID iD**: 0000-0002-8466-3316

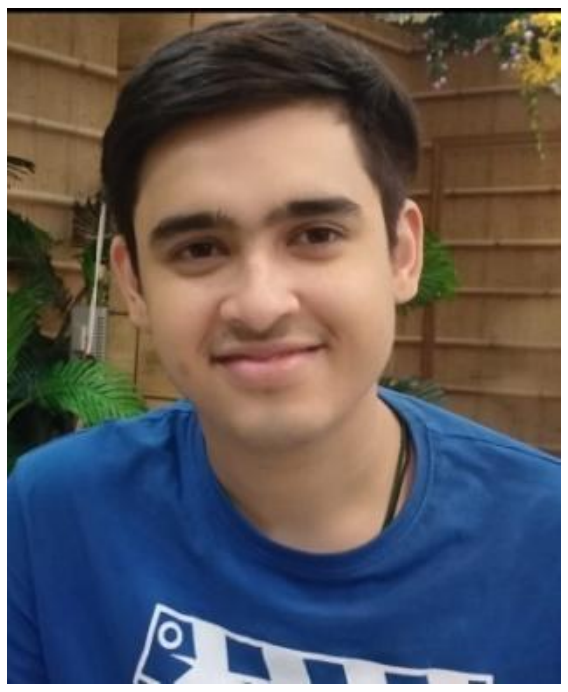

**Anubhav Sharma** is pursuing the Bachelor of Technology from LNMIIT, Jaipur, Rajasthan, India. His interests include research findings over new technologies and coding domains. His main research areas are Artificial Intelligence, Machine learning, language programming and developing projects.